\documentclass[acmsmall,nonacm,screen]{acmart}

\AtBeginDocument{%
  }

\setcopyright{none}

\usepackage[capitalise]{cleveref}
\crefname{section}{Sec.}{sections}
\Crefname{section}{Section}{Sections}
\crefname{lstlisting}{Lst.}{Lsts.}
\Crefname{lstlisting}{Listing}{Listings}

\usepackage{pgfplots}
\pgfplotsset{compat=1.18}
\usepackage{pgf-pie}
\usepackage{listings}
\usepackage{ltablex}
\usepackage{xparse}
\usepackage{hyperref}
\usepackage{subcaption}

\usepackage{enumitem}
\setlist[description]{leftmargin=10pt}

\newcommand{\edit}[1]{#1}

\usepackage{todonotes}

\usepackage{framed}
\newcounter{ObservationIdx}
\newenvironment{observation}%
{\begin{leftbar}
		\refstepcounter{ObservationIdx}
		\noindent\textsc{Observation\,\theObservationIdx.}\@ }%
	{\end{leftbar}}

\begin{document}
	\lstset{numbers=left,
		numbersep=5pt,
		numberstyle=\tiny,
		xleftmargin=\parindent,
		rulecolor=\color{black},
		frameround=tttt,
		frame=l}

\title{120 Domain-Specific Languages for Security}


\author{Markus Krausz}
\email{markus.krausz@rub.de}
\orcid{0000-0002-1362-423X}
\affiliation{%
  \institution{Ruhr University Bochum}
  \city{Bochum}
  \country{Germany}
}
\affiliation{%
  \institution{TÜV Informationstechnik GmbH}
  \city{Essen}
  \country{Germany}
}
\author{Sven Peldszus}
\email{sven.peldszus@rub.de}
\orcid{0000-0002-2604-0487}
\affiliation{%
  \institution{Ruhr University Bochum}
  \city{Bochum}
  \country{Germany}
}
\affiliation{%
	\institution{IT University of Copenhagen}
	\city{Copenhagen}
	\country{Denmark}
}
\author{Thorsten Berger}
\email{thorsten.berger@rub.de}
\orcid{0000-0002-3870-5167}
\affiliation{%
\institution{Ruhr University Bochum}
\city{Bochum}
\country{Germany}
}
\affiliation{%
\institution{Chalmers\,|\,University of Gothenburg}
\city{Gothenburg}
\country{Sweden}
}
\author{Francesco Regazzoni}
\email{f.regazzoni@uva.nl}
\orcid{0000-0001-6385-0780}
\affiliation{%
  \institution{University of Amsterdam}
  \city{Amsterdam}
  \country{The Netherlands}
}
\affiliation{%
  \institution{Universit\`{a} della Svizzera italiana (USI)}
  \city{Lugano}
  \country{Switzerland}
}
\author{Tim Güneysu}
\email{tim.gueneysu@rub.de}
\orcid{0000-0002-3293-4989}
\affiliation{%
  \institution{Ruhr University Bochum}
  \city{Bochum}
  \country{Germany}
}
\affiliation{%
  \institution{DFKI} 
  \city{Bremen}
  \country{Germany}
}

\renewcommand{\shortauthors}{Krausz et al.}

\begin{abstract} 
\looseness=-1
Security engineering---from creating security requirements to the implementation of security features, such as cryptography or authentification---is often supported by domain-specific languages (DSLs). While many security DSLs have been presented, a lack of overview and empirical data about these DSLs, such as which security aspects are addressed and when, hinders their effective use and further research. This systematic literature review examines 120 security DSLs regarding their security aspects and goals addressed, their language-specific characteristics, their integration into the software development lifecycle, and their evaluation. We observe \edit{a focus on individual development phases and} a high degree of fragmentation, which leads to opportunities for integration. The research community also needs to improve the usability and evaluation of security DSLs.
\end{abstract}




\maketitle

\section{Introduction} \label{sec:introdution}
\looseness=-1
Information security is a fundamental aspect when engineering software systems, as security breaches can have serious consequences.
Consider an Electronic Health Records (EHR) system, such as iTrust\,\cite{Heckman2018,Meneely}. It is crucial that sensitive patient information is not exposed to unauthorized entities, but it would be even worse if an attacker could manipulate medical records, potentially leading to incorrect and harmful treatments.
Although there is a relatively high level of awareness among developers about the security of software they develop\,\cite{Xie2011,Hermann2025icse}, successful attacks are still frequently reported, such as hospitals being forced to shut down due to attacks on their EHR systems\,\cite{Singh2016}.

\looseness=-1
A core challenge is that security spans the entire software development life cycle (SDLC)\,\cite{McGraw2004}.
Security has to be considered at different levels of abstraction while ensuring compliance among all developed artifacts\,\cite{Peldszus2022}.
However, already planning or realizing security on a single artifact is non-trivial and challenges developers\,\cite{Hermann2025icse}.
Relevant security requirements are often implicit\,\cite{Bartsch2011}, which makes addressing them hard.
Furthermore, common artifacts, such as source code, do not support easy reasoning about security. Security experts and other engineers should be provided with system representations that are tailored towards their tasks\,\cite{Peldszus2022}.

\looseness=-1
Contrary to programming languages, which are general-purpose, domain-specific languages (DSLs)\,\cite{dslbook}
are purpose-built and address the needs of a particular problem domain. DSLs can support the implementation of security features\,\cite{Peldszus2022,Hermann2025icse} as well as other phases, such as designing software systems. While DSLs are less expressive than general-purpose languages, they support developers and other engineers (e.g., security experts) by providing appropriate abstractions expressed in domain-specific terminology in one language\,\cite{dslbook}.
For instance, DSLs can represent software at a higher level of abstraction, removing irrelevant details, making them more intuitive and easier to understand. They can simplify the design or reasoning about specific aspects, such as security. 

\looseness=-1
This survey focuses on DSLs designed for information security, which aims at protecting information from unauthorized access, disclosure, modification, or destruction.
To counter the many possible attack vectors, a wide range of defensive techniques and building blocks exist at various system levels.
For example, access policies form a fundamental component, establishing guidelines and controls to determine who is granted access to specific resources. At a lower level, cryptography enforces security policies by preventing malicious parties from reading and manipulating sensitive data.

\looseness=-1
The scientific community has contributed numerous \emph{security DSLs}, which focus on specific security aspects and that offer security-specific language constructs. Such DSLs often target a particular phase of the SDLC. As we will show, the DSLs cover a broad spectrum of security, while research is highly fragmented. The lack of an overview and systematic empirical data on security DSLs limits the widespread adoption of security DSLs, hinders their development, and prevents future research.

This survey provides a comprehensive overview of security DSLs discussed in peer-reviewed academic literature.
Our goal is to identify patterns, trends, and open problems, and to suggest possible combinations of complementary security DSLs and integrations into the SDLC.
We organize the existing body of research on security DSLs, guided by the following research questions.
\vspace{-0.2cm}
\begin{enumerate}[wide, labelwidth=!, labelindent=0pt]
 	\item[\textbf{RQ1:}] \textit{What security DSLs have been presented in the scientific literature?}
	\noindent
	While researchers contributed many security DSLs, no systematic overview exists.
	We explore the existing literature and focus on DSLs that address security concerns in software development. 

	\item[\textbf{RQ2:}] \textit{What security aspects are targeted by security DSLs?}
	\noindent
	Security DSLs were designed to address different aspects of security, such as security goals, attacker models, and defense mechanisms.
	As such, they can be problem-oriented, focusing on general security planning and the identification of potential threats, or solution-oriented, facilitating the realization of  defense mechanisms.
	We capture these security aspects to contextualize the security DSLs.
	\vspace{2pt}

	\item[\textbf{RQ3:}] \textit{What phases of the development process are supported by security DSLs?}
\noindent\looseness=-1
	When developing secure software systems, security DSLs need to be integrated into the development process.
	DSLs can be relevant for a single or multiple development phases.
	We identify which SDLC phases are relevant for each security DSL, \edit{whether there are any phases particularly addressed by security DSLs,} and which security aspects the DSLs focus on in which phases.
	Specifically for DSLs that support multiple phases, we determine in which phase they are instantiated and in which they are used.
	\vspace{2pt}

	\item[\textbf{RQ4:}] \textit{What types of security DSLs exist, and how are they used?}
\noindent
	When designing a DSL, language engineers can choose between different types of DSLs, each type with its own characteristics.
	We examine these characteristics, such as whether a DSL is stand-alone (external) or embedded in an established language, such as a programming language, (internal) and whether a DSL is textual or visual.
	We also examine the tools associated with these DSLs to understand how they are used in practice.

	\item[\textbf{RQ5:}] \textit{What are the semantics of security DSLs?}
\noindent\looseness=-1
	Based on the targeted security goals and the targeted SDLC phases, we investigate how these are semantically realized, specifically what mechanism bring the DSL instances into effect.
	We capture how instances of security DSLs are executed or interpreted to achieve their security goals, or what kind of transformation is provided by the backend to propagate and utilize any security properties specified using the DSL.
	We also explore how security DSLs are integrated into the development process.
	\vspace{2pt}

	\item[\textbf{RQ6:}] \textit{How are security DSLs evaluated?}
\noindent
	Finally, we determine whether and how the DSLs have been evaluated, particularly their effectiveness.
	We analyze what security guarantees including properties like soundness, precision, and completeness were considered and perhaps even proven.
	We also identify and report quantitative and qualitative empirical data provided for the DSLs.
\end{enumerate}

\looseness=-1 
Our study supports developers, researchers, and tool vendors.
Developers and researchers obtain an overview of DSLs to support their work at different SDLC stages. They learn about examples, best practices, and universal solutions from existing security DSLs.
Researchers and tool vendors, such as those working on language-based security, possibly across multiple security domains
and attacker models, learn about security DSL concepts and about their integration, interaction, and combination.

\section{Background}
We introduce the necessary background on DSLs and information security, together with examples.

\subsection{Domain-Specific Languages}
\looseness=-1
A DSL is a programming, modeling, or specification language dedicated to a particular problem domain, problem representation, or solution\,\cite{Mernik2005,Fowler2011,dslbook}.
DSLs simplify the development process within a specific context or domain by providing abstractions and syntax that directly represent concepts and operations relevant to that domain.
DSLs typically hide unnecessary details and complexity present in general-purpose languages, allowing developers to focus on solving problems within their domain of expertise, such as software security.
Since DSLs are tailored to the needs of a domain, they are more intuitive, concise, and easier to use than general-purpose languages.
To this end, DSLs can improve productivity and code maintainability within specific domains by allowing developers to express solutions using terminology and abstractions familiar to domain experts.
However, designing and implementing DSLs requires careful consideration of the domain and the target users, making tradeoffs between expressiveness and ease of use.

\subsection{Information Security}
\looseness=-1
\emph{Information security} controls the flow of sensitive information at various system levels\,\cite{DBLP:books/daglib/0020262} with three core objectives: confidentiality, integrity, and availability\,\cite{Cawthra2020}.
Confidentiality guarantees that information is only accessible to intended parties.
Integrity prevents undesired modification of trusted data.
Availability prevents disruptions that could result in critical resources being unavailable.
Furthermore, privacy ensures confidentiality of private data\,\cite{Spiekermann2012};
authenticity ensures the integrity of messages' content and origin; accountability ensures availability and integrity of actors;
and non-repudiability ensures the accountability of messages' sender or receiver\,\cite{ISO27002}.

Before analyzing, enhancing, or building a system, the security \emph{requirements}
are usually determined and inferred from higher-level goals\,\cite{Tuerpe2017}.
For example, laws might demand privacy measures that necessitate controls at the enterprise level, which imply software requirements\,\cite{Deng2011}.

%
Requirements can also be identified after examining potential security threats.
\emph{Threat modeling}\,\cite{Shostack2008} is a method to comprehensively evaluate the security of a system by envisioning attack vectors and identifying gaps in the defense.
Threat modeling combined with the computation of probabilities enables proper assessment of security risks.
Attack patterns can also serve for detection during runtime.
%
After analyzing and planning, usually multiple defense mechanism are required to address the threats,
since attackers have many options and can target different system layers and components.
So, countermeasure are often accompanied by an \emph{attacker model}\,\cite{Rocchetto2016} that
defines the capabilities and limitations of potential attackers to defend against.

\emph{Access control}\,\cite{Sandhu1994} limits access to information, with foundations going back to the Bell-LaPadula\,\cite{bell1973secure} (confidentiality) and the Biba model\,\cite{biba1977integrity} (integrity).
Access control regulates physical or digital access to resources expressed as policies, which can be based on identities, roles, or attributes.
Multi-user operating systems, for example, manage read, write, and execute permissions for users.
\emph{Information flow control}\,\cite{Sabelfeld2003} is similar to access control, but at the level of program variables.
Information flow is often hard to track and protect in complex systems, and errors can lead to policy violations.
Flow control tracks data propagation in programs and verifies compliance with policies\,\cite{DBLP:journals/cacm/DenningD77}, which ensure integrity by prohibiting untrusted input to influence trusted output, or confidentiality by enforcing public output to be independent of secure input.

When transmitting sensitive data over insecure channels or storing it on unprotected devices, \emph{cryptography} with primitives (e.g., symmetric and asymmetric ciphers, or hash functions) maintains security.
With digital signatures\,\cite{DBLP:journals/iacr/CheonCDGHKLMSY23}, the integrity of messages can be verified.
A threat, however, are side-channel attacks, which use unintended information channels to extract sensitive information\,\cite{DBLP:conf/pkc/KrauszLRG23}. Examples are timing\,\cite{DBLP:conf/crypto/Kocher96} and power side-channel\,\cite{DBLP:conf/crypto/KocherJJ99} attacks.
In such attacks, adversaries do not attempt to break the cryptographic function, but exploit implementation weaknesses and unintended information leakage. For example, differences in execution time of a cryptographic function
can be exploited to infer secrets. 
Advanced cryptography includes concepts such as secure \emph{Multi-Party Computation (MPC)}\,\cite{goldreich1998secure}
and \emph{Fully Homomorphic Encryption (FHE)}\,\cite{DBLP:conf/stoc/Gentry09}.
MPC enables mutually distrustful parties to compute an arbitrary joint function of their private data, which is contributed by each party without revealing it to others. 
FHE even allows arbitrary computations on encrypted data without decryption, preserving confidentiality.

Besides preventive measures, security can be enhanced with \emph{intrusion detection} methods to identify attacks at runtime.
They classify\,\cite{axelsson2000intrusion} into anomaly and signature (pattern)-based approaches.
Prominent examples are network 
and host-based intrusion detection.
\section{Methodology}
\label{sec:method}
Following ACM SIGSOFT's Empirical Standards\,\cite{Ralph2021}, we used a systematic review methodology to study security DSLs.
In what follows, we describe our DSL selection criteria and the corresponding selection process, followed by the analysis methodology of the selected DSLs.

\subsection{DSL Selection}
We searched for security DSLs that can be used by developers, security professionals, and other stakeholders for developing secure software systems.
We focused on DSLs that were explicitly designed for security, specifically those that meet the following inclusion criteria

\begin{description}
	\item[\textbf{I1.}] The DSL can be used to address information security in software system engineering.
	\item[\textbf{I2.}] The DSL explicitly expresses one or more security aspects.
	\item[\textbf{I3.}] The DSL is presented in a peer-reviewed publication.
\end{description}

\begin{figure}[b]
	\vspace{-6pt}
	\centering
	\includegraphics[width=0.65\textwidth]{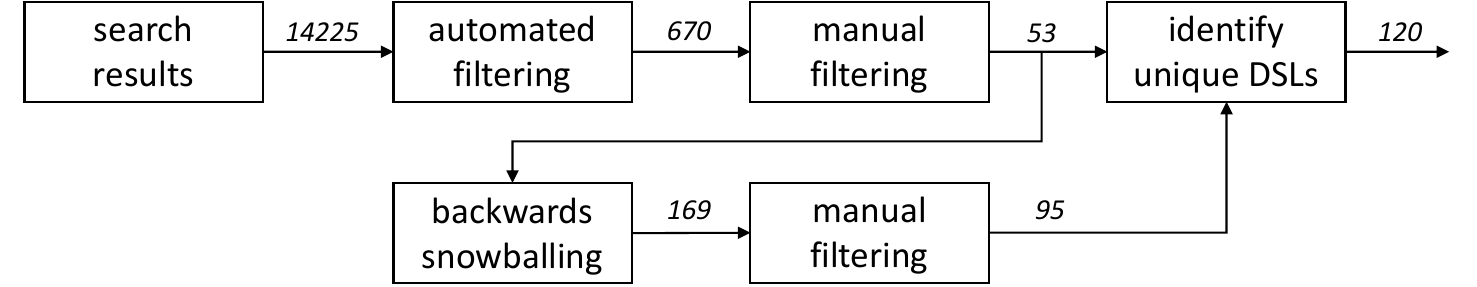}
	\vspace{-6pt}
	\caption{Literature filtering process.}
	\label{fig:sampling}
	\vspace{-.4cm}
\end{figure}

\looseness=-1
To systematically identify relevant DSLs from the academic literature (cf. I3) and to obtain a broad coverage of the landscape, we decided to use a generic search term in the initial search and then filter out irrelevant results.
Based on the initial set of publications identified, as recommended by ACM SIGSOFT's Empirical Standards\,\cite{Ralph2021}, we performed an iteration of backward snowballing\,\cite{Wohlin2014} to ensure that we did not miss relevant publications.
\Cref{fig:sampling} visualizes our sampling and filtering steps.
The first step (initial search) used the following search term on Semantic Scholar:
\textit{``security domain-specific language''}.
To avoid dated DSLs with low impact, to ensure the quality of the papers, and to prevent papers outside our scope, we defined six exclusion criteria.

\begin{description}
\item[\textbf{E1.}] We excluded papers published before 2010 in our initial search. However, we did not apply this criterion for snowballing, so relevant older papers are still in our dataset.
\item[\textbf{E2.}] We excluded papers not written in English.
\item[\textbf{E3.}] We excluded non-peer-reviewed papers as well as papers published by a predatory venue or publisher. Note that the security DSLs themselves do not have to be the contribution of scientific papers, but must have been discussed in at least one.
\item[\textbf{E4.}] We excluded papers that do not belong to the field of computer science.
\item[\textbf{E5.}] We excluded papers that do not focus on security DSLs and only mention them on the side.
\item[\textbf{E6.}] We excluded papers without a fully implemented DSL, but only theoretical concepts or models.
\end{description}

\looseness=-1
We searched Semantic Scholar using its public API and obtained 14,255 documents.
According to the exclusion criteria E1, E2, and E4, we limited the search to papers published 2010 or later, written in English, and the field of computer science.
Based on this set, we proceeded with automated filtering, first removing duplicates, which reduced the number of papers to 811.
From an initial manual inspection, we collected common keywords that indicate research irrelevant to us according to E5:

\begin{center}
	\textit{natural language, language model, language processing, security pattern,\\
	 machine translation, linguistic, speaker recognition, english language}
\end{center}

\looseness=-1
We then automatically scanned title, abstract, and paper for these keywords and reduced the papers to 670.
From these results, we continued with a manual selection based on the title and abstract.
We omitted remaining papers not written in English (E2).
Furthermore, we looked for evidence that the paper either presents a security DSL, discusses the usability or integration of an existing DSL, or surveys DSLs in a security domain, and excluded papers not meeting any of these three criteria (E5 and E6).
With 133 papers left, we narrowed down the set with a manual selection based on the full paper.
In doing so, we also discarded papers that were not published at a scientific conference or journal (e.g., three papers solely published as preprints).
We rejected three papers published at predatory journals according to Beall's list~\cite{beall} (E3).
Multiple publications were excluded simply because they did not cover a security DSL (E5), despite title and abstract indicating it.
We excluded several papers that presented type systems or formal calculi without a usable language (E6).
We also dismissed languages and tools that, although used in a security context, came without security-specific language features, but for example only a backend that provides security features.
\textsf{Prepose}\,\cite{DBLP:conf/sp/FigueiredoLMV16} is such an example, as the system clearly has a security goal, but the DSL is security agnostic.
For three papers, we could not access the full document and therefore omitted them.

\looseness=-1
After this filtering of the initial search results, we then added papers that were referenced as related work in our selected papers in one iteration of backwards snowballing.
This step brought up 169 new papers, for which we applied the same manual filtering,
leading to 95 additional relevant papers.
We ended up with 148 relevant papers covering 120 unique DSLs.

\subsection{DSL Analysis}
Our in-depth analysis of the identified security DSLs followed a four-step procedure.

\textit{1. Definition of aspects:}
For each research question, we defined relevant aspects according to which we analyzed the identified security DSLs.
We outlined these aspects already when discussing the research questions in \cref{sec:introdution} and provide detailed reasoning for each of them in \cref{sec:slr}.

\looseness=-1
\textit{2. Free-text responses:}
To describe each DSL according to these aspects, we first searched the corresponding papers from our dataset for relevant information.
If we did not find the information in the paper, or if the aspect was open-ended (e.g., whether there are practical applications of the DSL), we searched for further publications on the DSL and for websites or documentation of the DSL. \edit{In doing so, we did not limit ourselves to academic literature for analyzing the security DSLs.}
In this first step, we collected short free-text responses to each aspect.
To this end, we divided the selected papers between the first two authors of this paper.
While collecting the information on the security DSLs, the two authors were continuously synchronizing to discuss interpretations of single aspects for specific DSLs and to avoid divergence in the collected information.

\textit{3. Categorization:}
After collecting information for each DSL, we discussed all captured free texts and derived categories from them.
Since not all captured information was easily mappable to the identified categories, the categorization was followed by a data sanitation step in which the first two authors this time divided the individual aspects between them and went through all captured free text answers as well as the full papers again to assign every DSL, depending on the aspect, one or multiple of the identified categories.
Again, we supported this procedure with continuous discussions between the first and second author.

\looseness=-1
\textit{4. Statistical analysis:} Finally, after classifying all DSLs according to the categories identified for each aspect, we created statistics on the distribution of categories  per aspect, including correlations among them.
While discussed below, our supplementary material\,\cite{Krausz2025} provides further details. 
\section{Systematic Literature Review}
\label{sec:slr}
Our SLR identified 120 security DSLs. We now discuss the results of our in-depth analysis per RQ.

\begin{table}
	\caption{Security DSLs identified in the SLR (Part 1/2). \textbf{Categories}: Access Control (AC), Cryptography (CR), Information Flow Control (IFC), Intrusion Detection (ID), Requirements (R), Threat Modeling (TM); \textbf{SDLC Phases}: Requirements (R), Design (D), Code (C), Test (T), Deployment (DP), Runtime (RT)}
	\label{tab:papers}
	\vspace{-.6cm}
	\center
	\scriptsize
	\addtolength{\tabcolsep}{-0.4em}
	\begin{tabularx}{\textwidth}{llXlllll}
		\toprule
		\textbf{DSL} / \textit{\textbf{Author}} & \textbf{Year} & \textbf{Summary} & \textbf{Cat.} & \textbf{SDLC} & \textbf{References} & \textbf{Repo.}\\
		\midrule
		\textsf{ASL}					& 1997 & generic modeling of access control policies				& AC 		& D, C & \cite{218} & -- \\
		\textsf{Bro (Zeek)} 				& 1999 & detection of network attackers by passive monitoring		& ID 		& C, RT &\cite{260} & \cite{zeek} \\
		\textsf{AdeLe} 					& 2001 & model a database of known attack scenarios 				& TM, ID 	& C, RT & \cite{264} & -- \\
		\textsf{Ponder} 					& 2001 & generic modeling of access control policies				& AC 		& D, C & \cite{226} & -- \\
		\textsf{SPL} 					& 2001 & single description of global security policies				& AC 		& D, C & \cite{219} & -- \\
		\textsf{CORAS} 					& 2002 & risk assessment based on semiformal system model	 		& TM 		& R, D & \cite{300,301} & \cite{coras} \\
		\textsf{SecureUML}				& 2002 & modeling access control policies and their integration		& AC 		& D, C, RT & \cite{76,202,267,289} & \cite{secureuml} \\
		\textsf{STATL} 					& 2002 & intrusion detection of known attack scenarios				& ID		& C, RT & \cite{209} & -- \\
		\textsf{UMLsec} 					& 2002 & planning of secure system design							& R			& D & \cite{187, 347} & \cite{umlsec} \\
		\textsf{FlowCAML} 				& 2003 & secure information flow language							& IFC		& C & \cite{238} & \cite{flowcaml} \\
		\textsf{uniframe} 				& 2003 & security contracts for component descriptions				& AC		& D & \cite{291} & -- \\
		\textit{Ray et al.}		& 2003 & composition of access control frameworks					& AC		& D & \cite{270} & -- \\
		\textsf{Rei} 					& 2003 & policies applicable to pervasive environments	 			& AC		& C, RT & \cite{227, 228} & -- \\
		\textsf{XACML} 					& 2003 & common language for expressing security policies			& AC 		& C, RT & \cite{126, 229, 230} & \cite{XACML} \\
		\textsf{ESRML} 					& 2004 & enterprise information security requirements for IS0-17799	& R, AC 	& R & \cite{220} & -- \\
		\textsf{SAML}					& 2005 & exchanging authentication and authorization data 			& AC 		& C, RT & \cite{126} & \cite{355} \\
		\textsf{X-GTRBAC} 				& 2005 & expressive specification of RBAC							& AC 		& D & \cite{279, 280, 281, 285} & -- \\
		\textsf{WS-CoL} 					& 2005 & policy assertions of user requirements for web services 	& R			& C, RT & \cite{275} & -- \\
		\textsf{UMLintr}					& 2006 & specification of intrusion detection						& ID 		& R, D, C, RT & \cite{205} & -- \\
		\textsf{WS-SecurityPolicy}		& 2006 & security constraints and requirements for web services 	& R 		& C, RT & \cite{272} & \cite{274}\\
		\textsf{AsmLSec}					& 2007 & specifying attack scenarios								& ID 		& T & \cite{208} & -- \\
		\textsf{OSL} 					& 2007 & specify usage control policies								& AC 		& D, C & \cite{324} & -- \\
		\textsf{SECTET} 					& 2007 & policies for secure interorganizational workflows			& R, AC 	& R, D & \cite{244, 245, 246, 247, 290} & -- \\
		\textsf{SecureTROPOS} 			& 2007 & identification and planning of security requirements		& R			& R, D & \cite{210, 211, 212} & \cite{securetropos} \\
		\textsf{SMCL}					& 2007 & secure multi party computation 							& CR		& C & \cite{352} & -- \\
		\textit{Rodr\'{\i}guez et al.}& 2007 & embed security requirements into business processes	& R 		& R, D & \cite{257} & -- \\
		\textsf{WS-Policy4MASC} 			& 2007 & specification of monitoring and control policies			& R			& D, C & \cite{273} & -- \\
		\textsf{CA-RBAC} 				& 2008 & context-based access control								& AC 		& C, RT & \cite{286} & -- \\
		\textsf{SELinks/FABLE}			& 2008 & enforcement of security policies in web applications		& AC, IFC 	& C & \cite{83,200} & \cite{selinks}\\
		\textsf{SFDL 2.0}				& 2008 & secure multi party computation 							& CR		& C, RT & \cite{349,348} & \cite{sfdl}\\
		\textit{Breu et al.}	& 2008 & guide security investment decisions						& TM		& R, D & \cite{333} & -- \\
		\textsf{USE} 					& 2008 & specify and verify security policies						& AC		& D, C, RT & \cite{292} & -- \\
		\textsf{Cryptol} 				& 2009 & functional correctness of cryptography						& CR		& C & \cite{309} & \cite{cryptol} \\
		\textsf{Fabric} 					& 2009 & secure information flow in distributed systems				& IFC		& C & \cite{239,240} & \cite{fabric} \\
		\textsf{Nego-UCON\textsubscript{ABC}} 		& 2009 & access control for cloud services							& AC 		& D & \cite{306} & -- \\
		\textsf{ReAlSec} 				& 2009 & identify countermeasures for known attack patterns			& TM		& T, DP & \cite{214} & -- \\
		\textsf{S-Promela} 				& 2009 & generic security policies									& AC 		& C & \cite{215} & -- \\
		\textsf{Security Policy Model}	& 2009 & generating security policies from security requirements	& R, AC 	& R, D, C, RT & \cite{256, 259, 293} & -- \\
		\textsf{SPDL} 					& 2009 & simplifying the specification of SELinux security policies	& AC 		& C, RT & \cite{243} & \cite{seedit} \\
		\textsf{Pavlich-Mariscal et al.}& 2010 & access control policies							& AC 		& D, C & \cite{269} & -- \\
		\textsf{ADTrees}					& 2010 & specify and analyze attack-defense scenarios				& TM 		& D & \cite{340} & -- \\
		\textsf{ASLan++} 				& 2010 & security goals of distributed system designs				& AC 		& D & \cite{338} & -- \\
		\textsf{CapDL}					& 2010 & capability-based access control							& AC 		& D, RT & \cite{77} & \cite{capdl} \\
		\textsf{CRePE} 					& 2010 & fine-grained security policies for android applications	& AC 		& C, RT & \cite{250} & -- \\
		\textsf{SecPAL} 					& 2010 & simple but expressive, declarative authorization language	& AC 		& C & \cite{316} & -- \\
		\textsf{SSMF} 					& 2010 & identification of possible attack vectors					& TM 		& R & \cite{235} & -- \\
		\textsf{TASTYL}					& 2010 & secure multi party computation 							& CR 		& C, RT & \cite{353} & -- \\
		\textsf{X-Policy} 				& 2010 & dynamic execution permissions for collaborative systems	& AC 		& D & \cite{315} & -- \\
		\textsf{Caisson} 				& 2011 & secure information flow language							& IFC 		& C & \cite{236} & -- \\
		\textsf{cPLC} 					& 2011 & generate secure code for cryptographic two-party protocols	& CR 		& D, C & \cite{90} & -- \\
		\textit{Asnar et al.}		 		& 2011 & identification of permission-related threats			& R, TM 	& R & \cite{233} & -- \\
		\textsf{UACML} 					& 2011 & access control modeling language independent of model		& AC 		& D & \cite{71} & -- \\
		\textit{Bain et al.} 	& 2011 & high-level code for secure execution platforms (MPC \& FHE)& CR		& C & \cite{196} & -- \\
		\textsf{VALID} 					& 2011 & security goals for virtualized environments				& R 		& DP, RT & \cite{224} & -- \\
		\textsf{PTaCL} 					& 2012 & security policies based on attribute-based access control	& AC 		& D & \cite{346} & -- \\
		\textit{Hoisl et al.}	& 2012 & secure object flow in business process						& IFC 		& D & \cite{192} & -- \\
		\textsf{STRoBAC}  				& 2012 & spatial temporal role-based access control 				& AC 		& D, C, RT & \cite{283} & -- \\
		\textsf{TTCN-3} 					& 2012 & generation of penetration tests with attack patterns		& TM 		& T & \cite{213} & \cite{ttcn} \\
		\textit{Mitchell et al.}& 2012 & secure execution platform based on MPC or FHE				& CR		& C  & \cite{102} & -- \\
		\textsf{BCDSDL}					& 2013 & network availability with respect to botnet attacks		& IDE		& D & \cite{68} & -- \\
		\textsf{context-aware RBAC}		& 2013 & context-aware access control policies						& AC 		& D & \cite{193} & -- \\
		\textsf{CIL} 					& 2013 & specification of access control policies for Android		& AC 		& C, RT & \cite{25} & \cite{cil} \\
		\textsf{CySeMoL} 				& 2013 & probabilistic vulnerability assessment  					& TM 		& D, C & \cite{133, 136} & \cite{CySeMoL} \\
		\textsf{DJS}						& 2013 & solution for untrusted code in javascript					& CR 		& C & \cite{46} & -- \\
		\textsf{FAL} 					& 2013 & secure logging configurations								& CR 		& C, RT & \cite{222} & -- \\
		\textsf{Paragon} 				& 2013 & practical information flow control									& IFC 		& C & \cite{242} & \cite{paragon} \\
		\bottomrule
	\end{tabularx}
\end{table}
\begin{table}
	\caption{Security DSLs identified in the SLR (Part 2/2)}
	\label{tab:papers2}
	\vspace{-.6cm}
	\center
	\scriptsize
	\addtolength{\tabcolsep}{-0.4em}
	\begin{tabularx}{\textwidth}{llXllll}
		\toprule
		\textbf{DSL} / \textit{\textbf{Author}} & \textbf{Year} & \textbf{Summary} & \textbf{Cat.} & \textbf{SDLC}  & \textbf{References} & \textbf{Repo.} \\ 
		\midrule
		\textsf{RBAC DSL} 				& 2013 & specification of RBAC policies										& AC 	& D, C & \cite{104} & -- \\
		\textsf{SI* Asset \& Trust}		& 2013 & identify insider threats											& TM 	& R & \cite{105} & -- \\
		\textit{Ad{\~{a}}o et al.}& 2013 & secure key management											& CR 	& C & \cite{109} & -- \\
		\textsf{XFPM-RBAC} 				& 2013 & location-aware RABAC												& AC 	& D & \cite{166} & -- \\
		\textsf{ZQL} 					& 2013 & privacy-preserving data processing									& CR		& C & \cite{96} & -- \\
		\textsf{BreakGlassPolicies}		& 2014 & break-glass policies in role-based access control 					& AC 	& R, D & \cite{194} & -- \\
		\textsf{DelegationRBAC} 			& 2014 & definition of rights delegations									& AC 	& R, D & \cite{195} & -- \\
		\textsf{DPSL} 					& 2014 & facilitate specification of SOA security policies					& R 	& D, RT & \cite{169} & -- \\
		\textsf{Haka} 					& 2014 & describe protocols and apply security policies on network traffic	& ID 	& C, RT & \cite{266} & \cite{haka} \\
		\textsf{Jif}						& 2014 & practical information flow control									& IFC 	& C & \cite{188} & \cite{jif} \\
		\textsf{Sapper} 					& 2014 & timing-sensitive information flow control in hardware				& IFC 	& C & \cite{134} & -- \\
		\textsf{SecDSVL} 				& 2014 & specification of security requirements and threat modeling			& R, TM & R, D & \cite{11} & \cite{mdseatr} \\
		\textsf{Wysteria}				& 2014 & secure multi party computation										& CR 	& C, RT & \cite{37} & \cite{Wysteria}\\
		\textsf{XML Security DSL} 		& 2014 & simplifying the use of XML security mechanisms						& CR 	& C & \cite{175} & -- \\
		\textsf{AIDD} 					& 2015 & rule-based attack description and response language				& ID, TM& C & \cite{69} & -- \\
		\textsf{KEYRASE}					& 2015 & secure information erasure in software								& IFC, C& C & \cite{63} & -- \\
		\textsf{SecPDL}		 			& 2015 & distributed security policy self-management						& AC 	& D & \cite{158} & -- \\
		\textsf{PPLv2} 					& 2015 & access and usage control											& AC 	& C & \cite{118, 323} & -- \\
		\textsf{SecVerilog}				& 2015 & timing-sensitive information flow control in hardware				& IFC 	& C & \cite{132} & \cite{secverilog} \\
		\textsf{STS} 					& 2015 & modeling and reasoning about security requirements					& R 	& R, D & \cite{248} & \cite{sts} \\
		\textsf{SysML-Sec}				& 2015 & security requirements and identification of possible attacks		& R, TM & R, D, C & \cite{191} & -- \\
		\textsf{GemRBAC-DSL} 			& 2016 & expressive specification of RBAC									& AC 	& D, RT & \cite{225} & \cite{GemRBAC} \\
		\textsf{Lifty} 					& 2016 & automated repair for information flow control 						& IFC 	& C, RT & \cite{241} & -- \\
		\textsf{LJGS}					& 2016 & static information flow control supported by dynamic checks		& IFC 	& C & \cite{139} & \cite{ljgs} \\
		\textsf{pwnPr3d} 				& 2016 & attack graph generation											& TM 	& D & \cite{91} & \cite{PwnPr3d} \\
		\textsf{IoTsec} 					& 2017 & specification of security requirements								& R 	& R, D & \cite{206} & -- \\
		\textsf{Jasmin} 					& 2017 & efficient, safe, secure and correct cryptographic implementation	& CR 	& C & \cite{198} & \cite{jasmin} \\
		\textsf{Low*} 					& 2017 & efficient, safe, secure and correct cryptographic implementation	& CR 	& C & \cite{199} & \cite{fstar} \\
		\textsf{ProScript}				& 2017 & secure cryptographic protocol code									& CR 	& C & \cite{81} & \cite{proscript}\\
		\textsf{Vale}					& 2017 & secure and efficient cryptographic implementation					& CR 	& C & \cite{82} & \cite{vale}\\
		\textsf{SRESOFL} 				& 2017 & formal specification of security requirements						& R 	& R, D, C & \cite{5} & -- \\
		\textsf{STREAMS} 				& 2017 & specify information flow control policies for hardware				& IFC 	& C & \cite{43} & -- \\
		\textsf{ChiselFlow} 				& 2018 & timing-sensitive information flow control in hardware 				& IFC 	& C & \cite{299} & -- \\
		\textsf{IoT	Language}			& 2018 & threat modeling and risk assessment								& TM 	& D & \cite{146} & \cite{iot} \\
		\textsf{LOCKS} 					& 2018 & formal and convenient security goal specification					& R, TM & R, D, C & \cite{145} & -- \\
		\textsf{LUCON}					& 2018 & data-centric security policies for distributed systems				& IFC 	& C & \cite{101} & -- \\
		\textsf{MAL} 					& 2018 & threat modeling and attack simulation								& TM 	& D, T & \cite{17, 22, 189} & \cite{mal} \\
		\textsf{FaCT} 					& 2019 & timing side-channel secure cryptography							& CR 	& C & \cite{29,197} & \cite{fact} \\
		\textsf{Flamio} 					& 2019 & coarse-grained, dynamic, information-flow control					& IFC 	& C & \cite{88} & \cite{flamio}\\
		\textsf{IDS-DL} 					& 2019 & visual description of intrusion detection systems	 				& ID 	& D & \cite{28} & -- \\
		\textsf{UML-SR} 					& 2019 & specification of low-level security requirements					& R 	& R, D & \cite{2} & -- \\
		\textit{Cao et al.} 	& 2019 & access control between system components							& AC 	& D & \cite{18} & -- \\
		\textit{Zhao et al.}	& 2019 & integration of functional and security models						& R 	& D & \cite{182} & -- \\
		\textsf{QIF-RTL} 				& 2019 & quantitative information flow control in hardware					& IFC 	& C & \cite{12, 190} & \cite{qif-rtl} \\
		\textsf{Zee} 					& 2019 & information flow control for software systems with RTE				& IFC 	& C & \cite{21} & \cite{zee} \\
		\textsf{DynaMo} 					& 2020 & policy analysis and automated source code generation				& AC 	& C & \cite{157} & -- \\
		\textsf{Legal-GRL} 				& 2020 & legal compliance requirements										& R, AC & R, D, C, T & \cite{80} & -- \\
		\textsf{MAC\textsubscript{async}} 				& 2020 & secure information flow with respect to concurrency				& IFC 	& C & \cite{249} & -- \\
		\textsf{REAP} 					& 2020 & specify structure of security policies								& AC 	& D & \cite{157} & --\\
		\textsf{Co-Inflow} 				& 2021 & dynamic, coarse-grained information flow control					& IFC 	& C & \cite{54} & \cite{coinflow} \\
		\textsf{J\textsubscript{E}} 					& 2021 & language support for TEE integration					& IFC 	& C & \cite{6} & \cite{jelang} \\
		\textsf{SSEL} 					& 2021 & specification of secure SoC component communication				& R 	& R, D, C, T & \cite{1} & -- \\
		\textsf{Hamraz} 					& 2022 & information flow control in distributed system						& IFC 	& C & \cite{53} & -- \\
		\textsf{IFL (IFCIL)}			& 2022 & information flow control for SELinux								& IFC 	& C & \cite{13} & -- \\
		%
		%
		%
		\bottomrule
	\end{tabularx}
\end{table}

\subsection{RQ1: What Security DSLs Have Been Presented in the Scientific Literature?}
\looseness=-1
We first provide an overview of the security DSLs, their background, the scientific communities from which they originate, and how many of them are still actively maintained and accessible.
\Cref{tab:papers,tab:papers2} show the identified DSLs with their references.
Two DSLs share the same abbreviation, so we refer to the former\,\cite{158} as \textsf{SecPDL} and the latter\,\cite{243} as \textsf{SPDL}.
\Cref{fig:statistics:repo_update} shows that the DSLs were published between 1997 and 2022, and that the number of published DSLs increases steadily until a peak is reached in 2013.
We believe that DSLs have been published for most relevant security aspects---a hypothesis we explored in our detailed analysis discussed below.

\subsubsection{Background of the DSLs}
\looseness=-1
Most papers (49\,\%) are published in venues targeting the security community, followed by 11\,\% of venues in software engineering.
Other venues belong to the networks (9\,\%), applied computing (7\,\%), information systems (7\,\%), programming languages (7\,\%), and distributed computing (6\,\%) communities.
Note that a single venue may target multiple communities.
We could not assign a significant portion of the venues (3\,\%) to specific communities, because they target computer science in general.
In \cref{table:venues}, we list all conferences and journals where at least four of the papers we examined were published, as well as the assigned communities.

\begin{figure}
\begin{subfigure}{.45\textwidth}
	\begin{tikzpicture}
    \scriptsize
    \begin{axis}[
        xlabel=Number of Papers,
        xlabel near ticks,
        xtick distance=10,
        axis lines=left,
        axis line style={-},
        enlarge y limits=0.05,
        xmajorgrids=true,
        xmin=0,
        xmax=75,
        xbar,
        bar width=0.18cm,
        ytick=data,
        symbolic y coords={Security,Software Engineering,Network,Programming Languages,Information Systems,Applied Computing,Distributed Computing,Theoretical Computer Science,Operating Systems,Hardware,Requirements,Artificial Intelligence,Other},
        width=.7\textwidth,
        height=4.4cm,
    ]
    \addplot table [x=number, y=community, col sep=comma] {figures/venues.csv};
    \end{axis}
\end{tikzpicture}
\vspace{-6pt}
	\caption{Communities in which was published}
	\label{fig:statistics:communities}
\end{subfigure}
~
\begin{subfigure}{.55\textwidth}
\centering
\scriptsize
\renewcommand{\tabcolsep}{2pt}
\begin{tabularx}{\textwidth}{X>{\centering\arraybackslash}p{1.8cm}>{\centering\arraybackslash}p{.51cm}}
	\toprule
	\textbf{Venue Name} & \textbf{Community} & \hspace{-7pt}\textbf{Papers} \\
	\midrule
	Computer Security Foundations Symposium (CSF) & Security & 8\\
	Conference on Availability, Reliability and Security (ARES) & Security & 6\\
	Symposium on Security and Privacy (S\&P) & Security & 5\\
	USENIX Security Symposium & Security & 5\\
	Conference on Computer and Communications Security (CCS) & Security & 4\\
	Symposium on Policies for Distributed Systems and Networks (POLICY) & Distributed Computing, Network, Security & 4\\
	\bottomrule
\end{tabularx}
\caption{Venues with four or more papers}
\label{table:venues}
\end{subfigure}
\vspace{-20pt}
\caption{Communities of the publications from which the DSLs originate}
\vspace{-12pt}
\end{figure}

\begin{figure}[t]
	\begin{tikzpicture}
    \scriptsize
    \begin{axis}[
        ylabel=Number of DSLs,
        ytick distance=2,
        axis lines=left,
        axis line style={-},
        enlarge x limits=0.02,
        legend style={at={(0.2,0.9)},
        anchor=north, legend columns=0},
        xmajorgrids=false,
        xminorgrids=false,
        ymin=0,
        ymajorgrids=true,
        ybar=0pt,
        bar width=0.15cm,
        xtick=data,
        x tick label style={rotate=45,anchor=east,/pgf/number format/1000 sep=},
        width=\textwidth,
        height=3cm,
    ]
    \addplot table [x=year, y=published] {figures/annual.data};
    \addplot table [x=year, y=update] {figures/annual.data};
    \legend{published, last update}
    \end{axis}
\end{tikzpicture}
\vspace{-10pt}
	\caption{Publication year of the DSLs and latest update on their publicly available repository or website}
	\label{fig:statistics:repo_update}
	\vspace{-.6cm}
\end{figure}
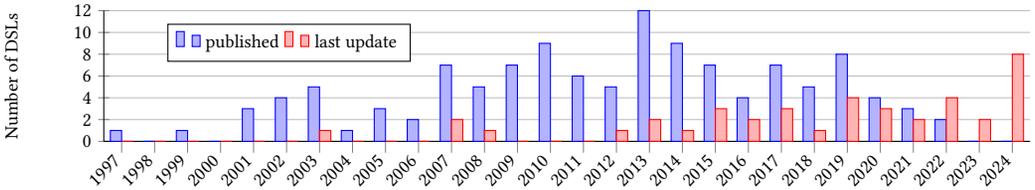

Only 40 of the 120 DSLs (33\,\%) have a publicly accessible repository or website.
Note that our initial search for publications covers the time span between 2010 and 2022, and with the backwards snowballing we further identified relevant papers dating back until 1997.
Many DSLs have been updated even after their release, as shown in \cref{fig:statistics:repo_update}.
In particular, at the beginning of 2025, 25\,\% of the updates happened recently in 2023 (2 DSLs) or 2024 (8 DSLs), 
indicating that quite a few of them are actively maintained and possibly also used in practice or academia.

\begin{observation}
	Security DSLs were primarily published in the mid-2010s, mainly from the security community, but also from other communities, such as software engineering.
While few DSLs are publicly available, many of them are still actively maintained.
\end{observation}

\subsection{RQ2: What Security Aspects are Targeted by the Security DSLs?}
\label{sec:rq2}
%
To obtain an over. over the security aspects covered by security DSLs, we analyze their fundamental security objectives and provide a coarse-grained categorization according to the security concept and the system scope targeted by each DSLs.
Furthermore, we capture the availability of attacker models and defense mechanisms.

First, we distinguished between problem- and solution-oriented DSLs.
Most languages (74\,\%) focus on solutions, e.g., by enhancing defense mechanism implementation and 23\,\% of the DSLs focus on (identifying) problems.
A few DSLs (3\,\%) are both, problem and solution oriented.

\subsubsection{Security Objectives}
\begin{figure}
	\begin{subfigure}{.49\textwidth}
		\begin{tikzpicture}
    \scriptsize
    \begin{axis}[
        xlabel=Number of DSLs,
        xtick distance=5,
        xlabel near ticks,
        axis lines=left,
        axis line style={-},
        enlarge y limits=0.08,
        xmajorgrids=true,
        xmin=0,
        xmax=60,
        xbar,
        bar width=0.15cm,
        ytick=data,
        symbolic y coords={Confidentiality,Integrity,Availability,Accountability,Authenticity,Nonrepudiability,Privacy,Any},
        width=0.9\textwidth,
        height=3.2cm,
    ]
    \addplot table [x=number, y=objective] {figures/objective.data};
    \end{axis}
\end{tikzpicture}
		\vspace{-16pt}
		\caption{Fundamental security objectives}
		\label{fig:statistics:objectives}
	\end{subfigure}
	~
	\begin{subfigure}{.49\textwidth}
		\begin{tikzpicture}
    \scriptsize
    \begin{axis}[
        xlabel=Number of DSLs,
        xtick distance=5,
        xlabel near ticks,
        axis lines=left,
        axis line style={-},
        enlarge y limits=0.10,
        xmajorgrids=true,
        xmin=0,
        xmax=45,
        xbar,
        bar width=0.18cm,
        ytick=data,
        symbolic y coords={Requirements,Access Control,Threat Modeling,Intrusion Detection,Information Flow Control,Cryptography},
        width=0.8\textwidth,
        height=3.2cm,
    ]
    \addplot table [x=number, y=category, col sep=comma] {figures/categories.csv};
    \end{axis}
\end{tikzpicture}
		\vspace{-16pt}
		\caption{Categories of security concerns}
		\label{fig:statistics:categories}
	\end{subfigure}
	\vspace{-8pt}
	\caption{Security aspects addressed by the DSLs}
	\vspace{-.4cm}
\end{figure}
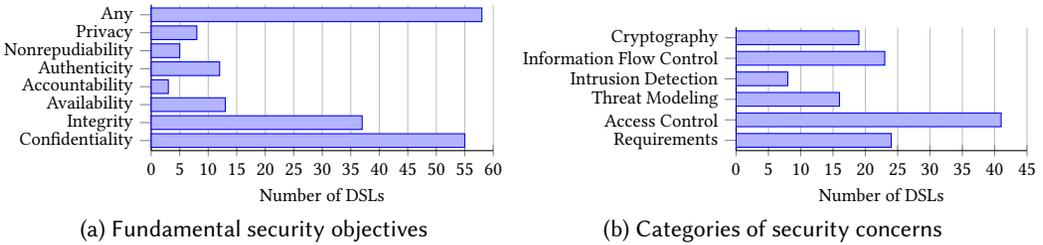

We investigated which security objectives the DSLs address and
were able to map the DSLs to seven fundamental security objectives (cf. \cref{fig:statistics:objectives}).

A DSL can address multiple security objectives, while some DSLs are not tied to a specific objective
and can be used for any potential security objective.
This can be the case, for example, for threat modeling languages.
However, the security objectives are explicitly formulated for only 46\,\% of the DSLs,
sometimes by making them part of the DSL or the framework around.
The other 54\,\% of the DSLs address security objectives implicitly,
e.g., in terms of considered~countermeasures.

Most DSLs target one of two core security objectives: confidentiality and integrity.
The former is addressed by 55 DSLs (46\,\%) and the latter by 37 DSLs (31\,\%).
Availability is only targeted by 13 DSLs (11\,\%)
and objectives like privacy are focused by even fewer languages.
However, 48\,\% of the DSLs do not consider a fixed set of objectives, but can potentially target any security objective.

For example, confidentiality is assured by languages for information flow control.
\Cref{lst:paragon} gives instances of illegal direct and indirect (via branching)
flow of secret information in the \textsf{Paragon} language.
It demonstrates how features like explicit declassification can be used to allow flow from secret to public values, which can be necessary in real-world applications.

\begin{lstlisting}[
	float,
	basicstyle=\scriptsize,
	label={lst:paragon},
	caption=An example in the \textsf{Paragon} language that demonstrates legal and illegal flow of data by direct and indirect flow including the feature of declassification
(from Broberg et al.~\protect\cite{242}).,
language=C,
belowskip=-1.5 \baselineskip]
void method(?Declass.high int highData) {
  Declass.low int lowData;
  lowData = highData; // Illegal direct flow
  lowData = Declass.declassify(highData); // OK
  if (highData == null) {lowData = null;} // Illegal indirect flow
}
\end{lstlisting}

\subsubsection{Categories of Security Concerns}
Furthermore, we investigated which security concerns the different DSLs address.
We identified six categories of targeted security concerns (see \cref{fig:statistics:categories}).
Thereby, a single DSL can target multiple of the identified categories.
\begin{description}
	\item[Requirements.]
	23 DSLs focus on security requirements.
	Some of them are tied to business process models, while others standardize the specification of requirements with policies, e.g., for web services.
	A typical example for this category is \textsf{SecureTROPOS}\,\cite{211}, a visual, graph-based DSL aimed
	at the identification and validation of security requirements for multiagent systems.
	\item[Access control.]
	Access control is the most dominant category we identified (43 DSLs).
	Multiple DSLs of this category focus on effective specification of access control policies.
	Furthermore, languages in this category are used to define access control models, such as
	the UML-based access control modeling language (\textsf{UACML})\,\cite{71}.
	\item[Threat modeling.]
	We found 17 DSLs that can be placed in the threat modeling category.
	Some of them facilitate the identification of attack vectors based on design models.
	Others allow efficient descriptions of attacks for simulations
	and to detect potential countermeasures.
	Part of this category is, e.g., \textsf{MAL}\,\cite{22}, a probabilistic meta threat modeling and attack simulation language.
	\item[Intrusion detection.]
	DSLs for intrusion detection systems (9 DSLs) are used to specify known attack patterns
	which can then be identified at runtime by the intrusion detection system and yield a response.
	Listing~\ref{lst:statl} demonstrates how STATL\,\cite{209} uses the fundamental concepts of states and transitions to express potentially vulnerable network protocol states for attack detection.
	\item[Information flow control.]
	Languages for information flow control (23 DSLs) are typically equip\-ped with a type system
	that features dedicated types to address confidentiality (secret vs. public)
	and integrity (trusted vs. untrusted).
	Via type inference, the flow of sensitive information over variables
	is automatically tracked, and the framework ensures dynamically or statically that, e.g.,
	secret variables do not reach public outputs.
	A well known representative is \textsf{Jif}\,\cite{188}.
	\item[Cryptography.]
	Cryptographic primitives are usually described with a mixture of plain text, pseudocode, and mathematical expressions, leading to ambiguous specifications.
	This category includes DSLs tackling this problem by including language constructs
	that enable an explicit and efficient expression of the underlying primitives (\textsf{Cryptol}\,\cite{309}).
	\Cref{lst:cryptol} shows how a part of the AES encryption can be specified with \textsf{Cryptol} using its built-in type for polynomials, primitive polynomial modulus, and multiplication operations.
	Other DSLs in this category enable a simple and correct usage of cryptography,
	or focus on side-channel secure implementations (\textsf{FaCT}\,\cite{197}).
	Some DSLs allow automated generation of MPC or FHE protocols from the specification of a function to be computed on the sensitive data (\textsf{Wysteria}\,\cite{37}), thus allowing non-cryptography experts to use this advanced cryptography.
	In total, 19 DSLs focus on cryptography.
\end{description}

\begin{figure}
\begin{minipage}{\textwidth}
 \begin{minipage}[t]{.47\textwidth}
\begin{lstlisting}[
	basicstyle=\scriptsize,label={lst:statl},
	caption={State transition in \textsf{STATL}
to identify half-open TCP connections (from Eckmann et al.~\protect\cite{209}).},
	keywords={transition,unwinding},
	belowskip=-1.8 \baselineskip,breaklines]
transition RST (s1 -> s0) unwinding {
  [IP ip [TCP tcp]] :
  (ip.header.src==victim_addr) &&
  (tcp.header.src==victim_port) &&
  (ip.header.dst==attacker_addr) &&
  (tcp.header.dst==attacker_port) &&
  (tcp.header.flags & TH_RST) }
\end{lstlisting}
\end{minipage}
\hspace{6pt}
 \begin{minipage}[t]{.47\textwidth}

\begin{lstlisting}[basicstyle=\scriptsize, label={lst:cryptol}, caption=Specification of multiplication in the Galois Field including the irreducible polynomial for AES written in \textsf{Cryptol} (from \textsf{Cryptol}~\protect\cite{309}),
	belowskip=-1.2 \baselineskip]
  poly = <|x^^8+x^^4+x^^3+x+1|>
  gf28Mult : (GF28, GF28) -> GF28
  gf28Mult (x, y) = pmod (pmult x y) poly
\end{lstlisting}
\end{minipage}
\end{minipage}
\end{figure}

Our categorization covers a wide range of subdomains within information security and indicates that DSLs have already been proposed and developed for most relevant security aspects.
It is noteworthy that most categories discussed above represent some type of defense mechanism, indicating that security DSLs focus on defense mechanisms.
In fact, 82\,\% of the DSLs support a concrete defense mechanism.
Many languages that are not associated with a specific defense mechanism aim to specify security requirements, and the choice of defense mechanisms is left to the user. Nevertheless, defenses play an important role in the use of these DSLs as well.

\subsubsection{System Scope}
Security concerns and defenses are applied considering different scopes of the information system.
For example, threat modeling can be performed using the same principles at a high level, where the entire enterprise is in scope, and at a lower level, where only the software is considered.
We used a coarse-grained categorization of information system scopes that reflects the common structuring of security research domains.
We mapped each DSL to a single scope that most closely matches the target scope of its security concept or countermeasure.

\begin{description}
	\item[Enterprise.]
	At the highest level, with 19 languages (16\,\%), the entire enterprise is in scope of the security concept of the DSL.
	Relevant aspects include business processes and human~interactions.
	\item[IT System.]
	We have assigned 73 DSLs (61\,\%) to the Information Technology (IT) system scope.
	This category contains combinations of multiple subsystems, e.g., via a network,
	but also the interaction of multiple programs, including operating systems.
	\item[Software.]
	20 DSLs (17\,\%) focus on the security of individual software, for example, software implementations of cryptographic primitives.
	\item[Hardware.]
	8 DSLs (7\,\%) focus on the security of individual hardware components.
	These are primarily (programming-language like) hardware description languages.
\end{description}

In summary, most DSLs facilitate security with a scope that considers multiple interacting components of the system. This is intuitively the case for DSLs supporting the requirements and design phases of secure information systems, but can also apply to programming languages with security mechanisms that have a scope beyond the software.

\subsubsection{Attacker Model}
Ideally, a defense mechanism is tied to an attacker model that defines the kind of attackers it aims to protect against.
This includes limitations of the attackers capabilities and allows determining if a countermeasure is sufficient for the applied scenario.

\begin{description}
	\item[Available.] Many DSLs (36\,\%) come with an explicitly specified attacker model.
	The categories cryptography and information flow control stand out with 87\,\% and 84\,\%, respectively.
	\item[Instance-specific.] For a number of DSLs (15\,\%) it is not possible to specify a general attacker model since it depends on the concrete subject system to which this DSL is applied. These DSLs are mostly design-time DSLs.
	\item[Focus of the DSL.] There are also some DSLs (7\,\% ) that do not come with an attacker model since it is their purpose to specify such an attacker model, e.g., as part of threat modeling.
	\item[None.] Unfortunately, with 51 DSLs (43\,\%), a significant number does not provide an attacker model. While for some DSLs it is not feasible to specify an attacker model, e.g., DSLs for reasoning about security requirements, these are only a few among the DSLs not providing an attacker model.
\end{description}

\begin{observation}
	Security DSLs exist for a wide range of security objectives, countermeasures, and system scopes. Unfortunately, a significant share lacks a precise attacker model.
\end{observation}

\subsection{RQ3: What Phases of the Development Process are Supported by the Security DSLs?}
To answer the third research question, we first investigated in which phases of the SDLC the security DSLs are used and what artifacts are expressed in them.
Thereafter, we investigated with which other artifacts they are integrated and how.

\subsubsection{Relevant Phases of the SDLC}
\begin{figure}[b]
\vspace{-.4cm}
	\begin{subfigure}{.53\columnwidth}
		\center
	\includegraphics[width=\textwidth]{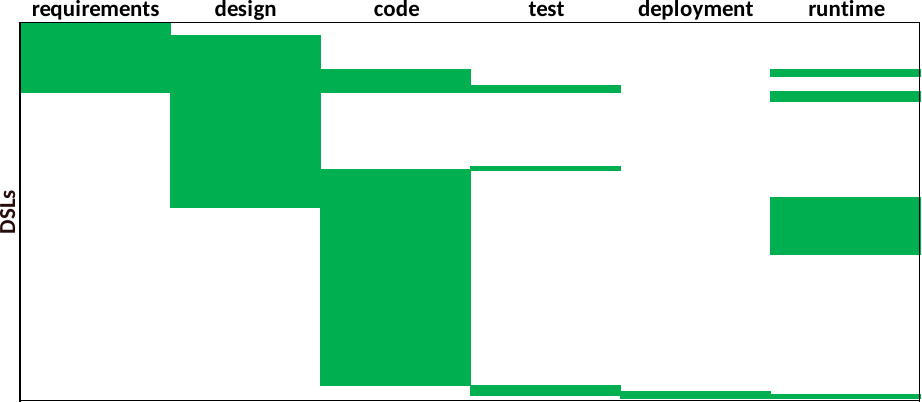}
	\caption{Use of the security DSLs across multiple phases}
	\label{fig:statistics:sdlc-aggregation}
	\end{subfigure}
	\hfill
	\begin{subfigure}{.44\columnwidth}
		\centering
	\begin{tikzpicture}
    \scriptsize
    \begin{axis}[
        ylabel=Number of DSLs,
        ytick distance=20,
        axis lines=left,
        axis line style={-},
        enlarge x limits=0.08,
        legend style={at={(0.7,0.9)}, anchor=north, legend columns=1},
        y label style={yshift=-0.2cm},
        xmajorgrids=false,
        xminorgrids=false,
        ymin=0,
        ymax=100,
        ymajorgrids=true,
        ybar=0pt,
        xtick=data,
        x tick label style={rotate=20,anchor=east,/pgf/number format/1000 sep=},
        symbolic x coords={Req.,Design,Code,Test,Depl.,Runtime},
        bar width=0.15cm,
        width=\textwidth,
        height=4.0cm,
    ]
    \addplot table [x=phase, y=instantiation, col sep=comma] {figures/phases.csv};
    \addplot table [x=phase, y=relevant, col sep=comma] {figures/phases.csv};
    \addplot table [x=phase, y=use, col sep=comma] {figures/phases.csv};
    \legend{instantiation, relevant, use}
    \end{axis}
\end{tikzpicture}
	\vspace{-8pt}
	\caption{DSL usage in the phases}
	\label{fig:statistics:sdlc-numbers}
\end{subfigure}
\vspace{-6pt}
\caption{SDLC phases relevant for using the security DSLs}
\label{fig:statistics:sdlc}
\vspace{-.4cm}
\end{figure}

The DSLs are used in six different phases of the SDLC, whereby 44\,\% of the DSLs are relevant in multiple phases.
On average, every DSL is used in 1.5 phases.
\Cref{fig:statistics:sdlc} shows how the DSLs distribute across the phases of the SDLC and concrete numbers of how often each phase is relevant for a DSL, in which phases instances of the DSLs are created, and in which phases these are actually put into use.

\begin{figure}[b]
	\vspace{-8pt}
	\includegraphics[width=.68\textwidth]{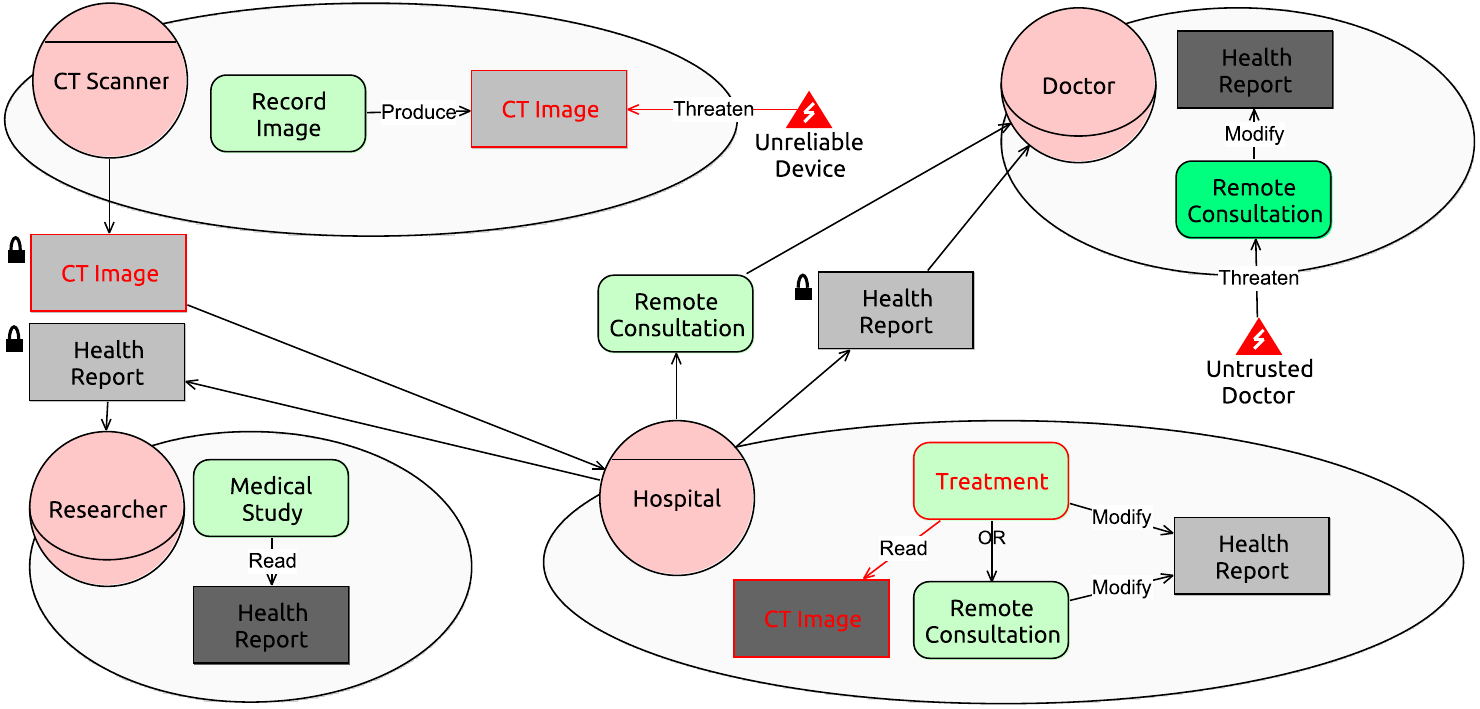}
	\vspace{-4pt}
	\caption{Partial actor diagram for an electronic health care system in \textsf{STS}}
	\label{fig:example:sts-scocial}
	\vspace{-8pt}
\end{figure}

\begin{description}
	\item[Requirements.] First,	for 22 DSLs (18\,\%) requirements engineering is a relevant phase.
	Usually, the DSLs support to identify and specify security requirements for the system to be developed.
	For example, \cref{fig:example:sts-scocial} shows how the requirements of a health care system are enriched with security requirements, i.e., confidentiality or integrity requirements on the data communicated, using \textsf{STS}.
	Requirements specified this way are reused in 82\,\% of the cases in later phases of the SDLC.

	\item[Design.] 49 DSLs (41\,\%) are instantiated in the design phase. Another six DSLs instantiated in the requirements phase are relevant to the design phase. In the design phase, more-detailed security requirements are specified. Here, a wide range of design models with different levels of abstraction are supported. For example, \textsf{UMLsec} allows to explicitly classify information in abstract domain models and allows planning detailed requirements on communication security, e.g., whether a connection needs to be encrypted(see \cref{fig:example:umlsec-deploy}). Design-level DSLs are mainly used in combination with the requirements phase and concrete code. Consequently, 40 DSLs are put into use in this phase. However, their use is also relevant in the test phase and at runtime.

	\item[Code.] With 76 DSLs, the majority of all DSLs (63\,\%) addresses the code phase on source code. 
	Of these DSLs, 67 DSLs are also instantiated in this phase.
	Code-only DSLs usually provide the means to implement secure code, e.g., containing security labels for secure data flow analysis, or for securely implementing cryptographic code.
	While most DSLs are only relevant to the code phase, there is still a significant number that also addresses design-time or runtime.
	Except for DSLs used at runtime, we observed no code-level DSL that is used in combination with any other phase unless it also covers the design phase.
	Mostly, concrete security objectives are planned at design-time, implemented or verified in code, thereby putting 54 of the relevant DSLs into active use in the implementation phase, and probably also enforced at runtime.

	\item[Test.] Only 6 DSLs (5\,\%) are used in the test phase, and all of them are used in further phases.
	However, the majority of these DSLs (4 DSLs) are instantiated in this phase as part of security tests, accompanying other parts of the DSL instantiated in earlier phases.
	The test phase is not the main focus of these DSLs, but they do allow, among other things, the generation of test code.

	\item[Deployment.] The fewest DSLs, only 2 DSLs (2\,\%), are used in the deployment phase.
	These DSLs are used in other phases, such as planning security goals for virtualized environments, and also support the actual secure deployment of the system.

	\item[Runtime.]	Finally, 52 DSLs are relevant to the runtime of the system, but often the DSL instance is not actively used, but is transformed into another format beforehand.
	For only 26 DSLs (22\,\%) the DSL instance is actively used at runtime.
	For DSLs at runtime, we can observe two different kinds of usage.
	First, security requirements specified at design-time being monitored at runtime, and second, concrete monitoring rules implemented in models or code being executed at runtime.
\end{description}

\edit{
	Security DSLs primarily focus on the code and design phases. During these phases, the majority of DSLs are intended to be instantiated in order to provide a concrete security benefit.
	Besides these two phases, security DSL instances are often used at runtime.
	Security DSLs seem to barely address the test and deployment phases.
}

\subsubsection{Artifacts Specified Using the DSLs}
\begin{figure}[b]
	\begin{tikzpicture}
		\scriptsize
		\def\dontprintone#1{\ifnum#1>1
		#1\,\%
		\fi}
		\pie[text=label, radius=1.45, rotate=352.8, before number=\dontprintone, after number=,
		color={blue!60, cyan!60, yellow!60, orange!60, red!60, blue!60!cyan!60, cyan!60!yellow!60,red!60!cyan!60, red!60!blue!60, orange!60!cyan!60, orange!60!magenta!60, lime!60}]
		{36/Policy, 27/Code, 13/Design Model, 9/Attack Description, 7/Requirements, 4/HW Description, 3/Intrusion Detection Rule, 2/Process Model, 1/Security Configuration}
	\end{tikzpicture}
	\vspace{-8pt}
	\caption{Kinds of artifacts expressed using the security DSLs}
	\label{fig:statistics:artifacts}
	\vspace{-.4cm}
\end{figure}
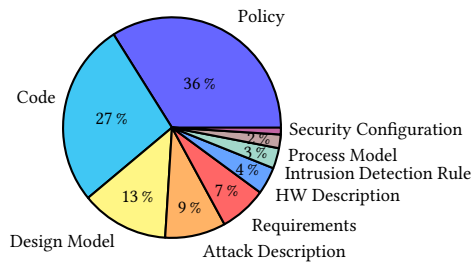

Following the development phases in which the DSLs are used, we observed a wide range of artifacts that are expressed using security DSLs (cf. \Cref{fig:statistics:artifacts}).

\begin{description}
	\item[Requirements.] Although 24 DSLs focus on security requirements, only 8 of them (7\,\% of all DSLs) are used to specify requirements.
	All other DSLs extend artifacts used in other phases with security requirements, such as in \textsf{UMLsec} (cf. \cref{fig:example:umlsec-deploy}) that enriches UML models with security requirements on communicated data and the security of communication paths used.

	\item[Process models.] Two DSLs express (business) process models with a visual representation of security considerations. While Rodr\'{\i}guez et al.\,\cite{257} extend the BPMN, the Security Policy Model\,\cite{256, 259, 293} defines its own process model notation.

	\item[Design models.] With 15 DSLs (13\,\%), many DSLs are used to create design models of the planned system.
	\cref{fig:example:umlsec-deploy} shows how \textsf{UMLsec} is used to plan secure deployments of software systems.

	\item[Code.] Corresponding to the SDLC phase of coding, where most DSLs are used, with 32 DSLs (27\,\%), code is the second-largest artifact category.
	DSLs are used to create custom code artifacts, such as \textsf{STATL} (see \cref{lst:statl}), as well as code artifacts in common general purpose languages, such as \textsf{TASTYL} which is based on python. We recall that only 17\,\% of the DSLs facilitate security with a scope limited to software, for example, by addressing side-channel security or memory-safety. Therefore, many DSLs with code artifacts feature defense mechanisms that consider the interaction of several system components, e.g., distributed systems.

	\item[Hardware descriptions.] Five DSLs (4\,\%) are hardware description languages that express security considerations of hardware, typically in the form of logic circuits that can be generated and come with some security guarantees.

	\item[Attack/defense descriptions.]	Eleven DSLs (9\,\%) are used to describe attacks and some of them also appropriate defenses. Most of these DSLs use a graph-based representation such as attack trees\,\cite{340} or probabilistic attack-defense graphs\,\cite{17, 22, 189}.

	\item[Policy files.] With (43 DSLs, 36\,\%), policy files are the largest artifact category we encountered.
	Since policies are a fundamental aspect of security and a component of various defense mechanisms, especially access control systems, which were the most common category of security aspects among all DSLs (see \cref{sec:rq2}), this is an expected observation.
	Within this category, 29 DSLs are used to specify plain policies, while 13 DSLs (11\,\%) are used to create concrete access control models that go beyond simple policies.
	One DSL (\textsf{PPLv2}\,\cite{118}) specifies detailed models for access and usage control, which even goes beyond what can be expressed in the policy files or access control models considered above.

	\item[Intrusion detection rules.] Three DSLs (3\,\%) define some form of intrusion detection rules.
	\textsf{AIDD} \cite{69} and \textsf{IDS-DL}\,\cite{28} facilitate the specification of intrusion detection rules, which could be seen as special policies or attack and defense descriptions.
	\textsf{Haka}\,\cite{266} allows specifying communication protocols in combination with filtering rules to identify and address malware network activities.

	\item[Security configuration.] Lastly, \textsf{FAL}\,\cite{222} is used to specify configuration files of a secure logger.
\end{description}

\edit{
	Following the low coverage of the test and deployment phases, the artifacts expressed using DSLs do not cover test cases or concrete deployment artifacts.
	Security configurations are the closest related artifacts to deployment, but only one DSL considers them.
	Test-specific artifacts that are explicitly expressed using a DSL are entirely missing.
}

\subsubsection{Integration With Other Artifacts}
The investigated DSLs are usually not used in isolation but 62\,\% of all DSLs are integrated with other artifacts.
We consider an DSL as being integrated with an artifact if it has an explicit relation to this artifact, e.g., by extending it or explicitly referring to it.

\begin{figure}[b]
\vspace{-.2cm}
	\includegraphics[width=.7\textwidth]{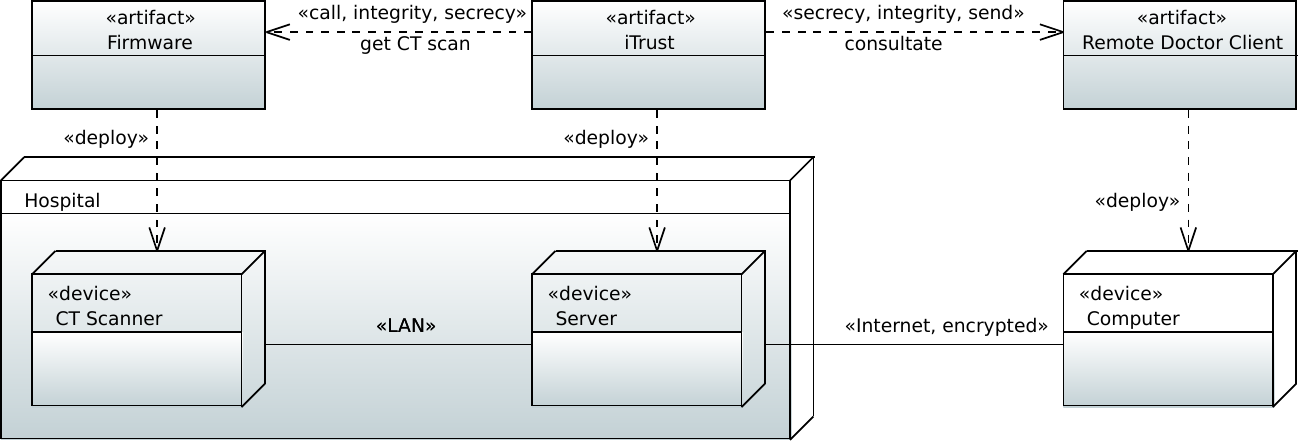}
		\vspace{-4pt}
	\caption{UML deployment diagram of a health records system annotated with \textsf{UMLsec} for planning security of communication paths based on labeled information communicated over them (oriented on Peldszus\,\cite{Peldszus2022})}
	\label{fig:example:umlsec-deploy}
\end{figure}

\begin{description}
	\item[Requirements.] Four DSLs	(3\,\%) enrich requirements expressed in different artifacts with security-related aspects such as relevant threats for concrete requirements.

	\item[Design models.]	Similarly, 20 DSLs (17\,\%) enrich design models with security objectives, threats, and planned features.
	For example, in \Cref{fig:example:umlsec-deploy}, \textsf{UMLsec} is used to enrich a UML deployment diagram with information on data sensibility and the security requirements of its communication.

	\item[Attack/defense descriptions.]	Only the \textsf{LOCKS} DSL is integrated with attack descriptions in SAM models for specifying security goals.

	\item[Code.] Most DSLs (33 DSLs, 28\,\%) are integrated with some form of code.
	One half of these DSLs is complied into code in some programming language that will be embedded into functional source code and the other half directly extends functional source code.

	\item[Policies.] Thirteen DSLs (11\,\%) are integrated with policies, thereof, nine DSLs express themselves policies or access/usage control models.
	Usually, these are integrated or compatible with a policy standard, such as \textsf{XACML} (6 DSLs).

	\item[Runtime components.] Nine percent of the DSLs are integrated with runtime components. Eight DSLs are integrated with components of the system at runtime, usually meaning that the instance of the DSL is executed at runtime in combination with this runtime component, e.g., compiled functional source code.
	Two DSLs are integrated with intrusion detection systems.

	\item[Hardware.] Only \textsf{STREAMS}\,\cite{43} integrates with existing hardware description artifacts by providing a hardware module for use in Verilog.
	All other DSLs that target hardware description specify these in their own languages, but do not integrate with existing specifications.

\end{description}

Unlike the significant number of DSLs that are used in multiple SDLC phases would make one expect, only 7 DSLs are integrated with multiple artifacts.
This is due to the fact that we consider a transformation of a DSL instance into another format not as an actual integration a developer has to deal with.
DSLs that are integrated with multiple artifacts are usually design models that are extended or used to express requirements or policies.

\subsubsection{Kind of Integration} Fewer than half of the DSLs (51 DSLs) are relevant to multiple development phases. We identified five variants in which DSLs can be integrated into the development. A single DSL can be integrated using multiple of these variants.

\begin{description}
	\item[Planning.] Most of the DSLs (23\,\%) serve as languages for planning artifacts.
	The use of DSLs enables the documentation of security requirements, planned countermeasures, and other security aspects that must be considered or realized in subsequent phases of the SDLC. These serve as requirements for subsequent phases, with the manual usage of artifacts created using the corresponding DSLs representing one of their primary purposes.

	\item[Enriching artifacts.] The second-highest number of DSLs (18\,\%) is integrated into the SDLC by enriching artifacts that are created as part of the development process anyway, e.g., for planning or documentation purposes, with security considerations.

	\item[Detailing.] A significant proportion of DSLs (11\,\%) permit the continuous detailing of the security aspects expressed using them over the phases of the SDLC. As the development of the system progresses, the DSL instances are enhanced with further information, particularly in relation to access control policies, security requirements, and threat models.

	\item[Reuse.] Many artifacts created using DSLs (13\,\%) are reused in later phases of the SDLC by other security engineering techniques, such as access control policies derived at design time that serve as input to an access control engine at runtime.

	\item[Traceability.] Finally, the Access Control Framework\,\cite{269} also allows tracing security requirements across different artifacts created in the SDLC.

	\item[None.] Most DSLs (58\,\%) have been designed with a single artifact and are used in isolation.
\end{description}

\begin{observation}
	Security DSLs are used in all phases of the SDLC, but 58\,\% of them do not feature any means of integration into the SDLC.
	There is a significant focus on design-time and code-level DSLs, while deployment and testing are underrepresented.
	The most common artifacts are security policies, code, or design models.
	Primarily DSLs targeting the requirements engineering or design phase are used in multiple SDLC phases.
\end{observation}

\subsection{RQ4: What Types of Security DSLs Exist, and How are They Used?}
\label{sec:rq4}
For embedding security aspects into development processes as discussed above, DSLs can be specified following various language design principles.
To get a better understanding of the technical perspective on security DSLs, first, we look at how the DSLs are defined and what are their characteristics.
%
 To this end, we discuss whether the DSLs are an extension of an existing language or are newly designed languages and how instances are represented.
 Thereafter, we discuss how developers work with the DSLs and how they are supported.

\subsubsection{Language Design}
When examining the artifacts created using the DSLs, we observed that some are defined entirely by the DSL's syntax, while others are enriched versions of other languages.
\begin{description}
	\item[External DSLs.] Most commonly (66\,\%), we observed external DSLs that come with custom-tailored parsers or interpreters specifically designed for the language features of the DSL.

	\item[Internal DSLs.] Internal DSLs (34\,\%) enrich an existing language and use its language infrastructure.
	The most common host language we observed is the Unified Modeling Language (UML), which is extended by 40\,\% of all internal DSLs via so-called UML profiles.
	In total, 50\,\% of internal DSLs extend modeling languages, such as SysML, BPMN, and AADL.
	The second-largest group of internal DSLs extend programming languages (35\,\%).
	Thereby, Java and Haskel are the most extended languages (4 DSLs each).
	Finally, three DSLs each enrich policy languages (IF, WS-Policy, and XCAML) and requirements languages (GRL and two times SI*).
	For example, Asnar et al.\,\cite{233} extend the SI* goal modeling language\,\cite{Massacci2010} with threat modeling capabilities.
\end{description}

\subsubsection{Relation to Existing DSLs}
The DSLs analyzed are usually new languages, which in the case of internal DSLs may be related to an underlying modeling or programming language.
However, 18\,\% of them are based on an existing DSL and extend it with additional aspects.
Extended DSLs are mainly DSLs for specifying access control policies (31\,\%), e.g., \textsf{WS-CoL} and \textsf{WS-Policy4MASC} extend the WS-Policy DSL with additional policy concepts.
Secure code DSLs are also being extended, e.g., \textsf{Fabric} and \textsf{$J_E$} both extend \textsf{Jif} with additional concepts.
In addition, the percentage of threat modeling DSLs that extend another security-specific language is remarkably high at 35\,\%.

\subsubsection{Instance Representation}
The second characteristic of a language is how instances are represented and made accessible to users.
This can be entirely textual, visual, or a combination.

\begin{description}
	\item[Textual DSLs.] The majority of DSLs (69\,\%) are textual languages. The DSLs define a textual syntax for the language, allowing users to specify instances in any text editor.
	DSLs such as \textsf{Jif} extend the language syntax of Java with additional elements and can be used in a similar way to standard Java source code.
	\textsf{STARTL}, \textsf{Paragon}, \textsf{Cryptol}, and many others define their own programming language-like syntax.
    Policy languages often use an XML syntax.
	Almost all DSLs in the information flow control and cryptography categories are textual languages.

	\item[Visual DSLs.] \textsf{STS} and \textsf{UMLsec} are examples for a visual DSLs.
	\Cref{fig:example:sts-scocial} provides a partial example of a social diagram in \textsf{STS}.
	The visual notation of \textsf{STS} is oriented on SI* goal models, while \textsf{UMLsec} extends the UML.
	In the back, both models are serialized as XML files, but these are not intended for manual use.
	Overall, 24\,\% of the DSLs we examined work on a visual representation.
	Interestingly, 64\,\%  of problem-oriented DSLs are visual, and 11\,\% are textual and visual.
	\item[Textual \& Visual DSLs.] A few DSLs (7\,\%) allow users to work on both kinds of representations. Depending on the concrete task, they can select the more suitable representation of the instance, i.e., users are often faster in writing instances in the textual representation but are better in understanding and debugging them in presented visually.
	
\end{description}

\subsubsection{Editing Support} 
\begin{figure}
	\begin{tikzpicture}
		\scriptsize
		\pie[text=label, radius=.8, rotate=270]{64/None, 16/Basic, 8/Extended, 13/Sophisticated}
		\pie[text=label, radius=.8, pos={4.5,0}, rotate=270]{69/Textual, 24/Visual, 7/Text. \& Vis.}
		\pie[text=label, radius=.8, pos={9.0,0}, rotate=270]{66/External, 34/Internal}
	\end{tikzpicture}
	\vspace{-8pt}
	\caption{Statistics on editing support and instance representation and language design (from left to right)}
	\label{fig:statistics:editingsupport}
	\vspace{-4pt}
\end{figure}
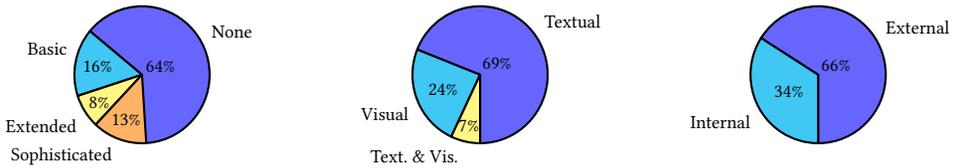

To allow users to work efficiently and effectively with DSLs it is important to provide them with editing support\,\cite{Barros2022}.
Editing support ranges from simple batch syntax checking, such as usually provided by compilers, over its embedding into IDEs for live feedback on the syntax, and syntax highlighting to semantic feedback on the expressed aspects or suggestions for auto-completion.
Among the DSLs, we identified four degrees of editing support (cf.~\cref{fig:statistics:editingsupport}).

\begin{description}
	\item[None.] Most DSLs come with no editing support at all (64\,\%). Users have to write instances with any editor and are not supported in doing so. 

	\item[Basic.] At least some basic editing support, usually comprising syntax highlighting and checking, is provided for 16\,\% of the DSLs.

	\item[Extended.] A few DSLs (8\,\%) extend basic editing support with tools that facilitate the specification of language instances. This support usually, includes simple features such as the completion of keywords defined in the language, but does not go as far as considering the expressed semantics.

	\item[Sophisticated.] The highest degree of editing support comprises instance-specific checks of the expressed semantics and sophisticated features such as auto-completion and is provided by 13\,\% of the DSLs. This degree is mainly implemented via language-engineering frameworks, such as Xtext\,\cite{Bettini2016} that automatically generate editing support features\,\cite{Barros2022} or UML-based languages that are enriched with semantic checks embedded via OCL constraints.
\end{description}

\begin{observation}
	Security DSLs in general are usually textual, but most problem-oriented DSLs have a (partially) visual representation.
	Editing support is only available for 37\,\% of the DSLs.
\end{observation}

\subsection{RQ5: What are the Semantics of the Security DSLs?}

To better understand how the security DSLs affect the development of secure software systems, we investigated their semantic usage.

\subsubsection{Usage Purpose} 
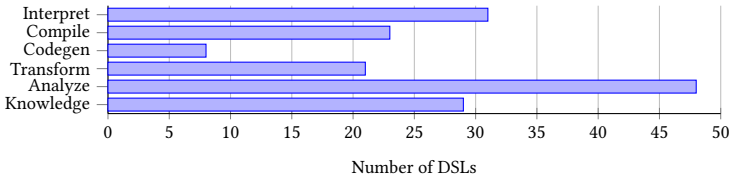
\begin{figure}
	    \begin{tikzpicture}
        \scriptsize
        \begin{axis}[
            xlabel=Number of DSLs,
            xtick distance=5,
            axis lines=left,
            axis line style={-},
            enlarge y limits=0.10,
            xmajorgrids=true,
            xmin=0,
            xmax=50,
            xbar,
            bar width=0.17cm,
            ytick=data,
            symbolic y coords={Knowledge,Analyze,Transform,Codegen,Compile,Interpret},
            width=0.7\textwidth,
            height=3cm,
        ]
        \addplot table [x=number, y=semantic, col sep=comma] {figures/usage.csv};
        \end{axis}
    \end{tikzpicture}
	\vspace{-8pt}
	\caption{Semantic usage of the DSLs (one DSL can be used in multiple ways)}
	\label{fig:statistics:usage}
	\vspace{-10pt}
\end{figure}

First, we looked at the semantic purpose for which the security DSLs are used in the development of secure software systems (see \cref{fig:statistics:usage}). A single DSL can be used in multiple semantic ways and on average the DSLs are intended to be used in 1.5 ways.

\begin{description}
	\item[Knowledge.] With 29 DSLs (24\,\%), a significant number of DSLs serve as a knowledge exchange format between stakeholders for documentation purposes. For example, the security requirements specified in the DSLs are given to the developers, who then implement them.

	\item[Analyze.] 52 DSLs (43\,\%) provide support to developers by analyzing the system as represented by an instance for security issues, such as identifying insecure design decisions.

	\item[Transform.] There are 24 DSLs (20\,\%) that are intended to be the source of an automated transformation into another instance format, i.e., another DSL.
	Typically, these DSLs allow for a simpler specification of the security aspects expressed than the usually more generic target format and domain-specific knowledge that is hidden by the DSL is added through the transformation.
	In most cases, access control policies derived at design time are transformed into other formats that can be executed at runtime, e.g., by an access control engine.

	\item[Codegen.] The generation of secure source or byte code in a general purpose programming language that then can be embedded by developers into the implementation of the system is supported by 9 DSLs (8\,\%).
	The code generated this way is usually not executable on its own but must be manually embedded into the implementation of the software system.

	\item[Compile.] 20 DSLs (17\,\%) compile an instance into an executable format.

	\item[Interpret.] 33 DSLs (28\,\%) are executed by an interpreter.
	DSLs used in this way mainly cover security constraints such as access control policies that are enforced by a runtime engine.
\end{description}

\subsubsection{Automation}
DSLs are commonly provided with tooling that automates certain aspects of their semantic usage.
Among the DSLs that come with such tooling, the average number of automated semantic usages is 1.2, which is much lower than the intended average of 1.5 usages.

\begin{description}
	\item[None.] While capturing knowledge for the use by humans is the only semantic usage purpose of DSLs that cannot be automated and only 15 DSLs have solely this purpose, 33 DSLs (28\,\%) do not come with any automation.
	Tough, some automation is often outlined as future work.

	\item[Analyze.] The majority of DSLs (36\,\%) come with an automation of their proposed analyses, e.g., for identifying security issues.
	 While 52 DSLs propose analysis as semantic usage, with 43 DSLs most of them actually automate this usage (83\,\%).
	Analyzes are mainly hard-coded, but are also frequently implemented using query or constraint languages such as the OCL.
	\edit{While automated analyses mostly do not provide guaranteed security aspects, they help in identifying vulnerabilities and increase the trust in the artifacts expressed using the DSL.}

	\item[Transform.] All DSLs that propose a transformation, with the exception of four (83\,\% of the DSLs that propose transformations), automate this transformation.
	In total, 20 DSLs (17\,\%) automate transformations. Transformations are mostly implemented in general purpose languages and model transformation languages build the exception.

	\item[Codegen.] All nine DSLs that propose code generation actually implement tooling for generating source code from the DSL instances. \edit{This way, vulnerabilities can be avoided and the security of the system can be increased.}

	\item[Compile.] Only 9 of the 20 DSLs that propose to compilation actually implement a compiler (60\,\%).

	\item[Interpret.] Similarly, only 64\,\% of the 20 DSLs that propose interpretation also realize it.
\end{description}

83\,\% of the DSLs come with tooling for automated vulnerability detection or for integration with other SDLC phases through transformation.
However, putting the DSLs actually into action remains mainly theoretical and is often not implemented. One exception is code generation for usage by developers as part of their implementations which is implemented for all 9 DSLs for which such a code generation has been proposed.

\begin{observation}
	The common semantic purposes of security DSLs are analysis, knowledge exchange, or (either direct or via indirect means like transformations or code generation) code execution. Automation, in particular by compilation and interpretation is often not implemented.
\end{observation}

\subsection{How are the Security DSLs Evaluated?}
\label{sec:rq6-eval}
Our final research question focuses on the evaluation of the DSLs, \edit{in particular the evaluation methodology and objective.}
Three DSLs do not come with any evaluation, but one of them is an open standard and widely used in practice. Therefore, we also investigated whether the DSLs have been applied in practice, e.g., as part of a case study.

\subsubsection{Empirical Methods}

Empirical evaluations assess the claims associated with the DSLs by gathering and analyzing observable evidence or data. Claims can relate to the security goal (e.g., effectiveness to prevent security issues) or quality aspects (e.g., usability). To obtain an overview, we first determined using which empirical method the DSLs and their supporting tooling are evaluated.

\begin{description}
	\item[None.] Unfortunately, most DSLs (44 DSLs) come without any empirical evaluation.

	\looseness=-1
	\item[Example.] 29 DSLs are illustrated on smaller examples, which are often realistic, but still artificial.
	However, only for 17 these examples are actually part of an empirical evaluation.
	In these cases, the examples make up for a benchmark suite or are used in experiments with users of the DSL.
	In the other 12 DSLs, the examples are mainly a vehicle to illustratively explain the DSL to the reader but are sold as an empirical evaluation.

	\item[Experiment.] Only for 16 DSLs their authors performed systematic experiments to evaluate the DSL, mainly concerning the aspect of performance (13 DSLs).
	Besides this, for 3 DSLs scalability is evaluated based on experiments and the effectiveness of 2 DSLs.

	\item[Case study.] The main vehicle for empirical evaluation is the case study (42 DSLs).
	Thereby, the main objective of a case study is to assess the feasibility of the DSL for the intended use case in its natural context\,\cite{Runeson2009}.
	To this end, the DSL is applied to one or more case studies and the application (and experiences) are reported in the paper.
	On average, the papers in this category contain 2.5 case studies, with a maximum of 12 case studies, resulting in a median of 1 case study.
	The case studies themselves are real-world systems or realistic systems of varying sizes.
	External experts are involved only~rarely.

	\item[User study.] Seven user studies were conducted to evaluate the concrete usage of the DSL by target stakeholders or in the concrete application context. The studies typically aim to assess the usability for stakeholders (4 DSLs), the applicability to real-world contexts (4 DSLs), and the understandability (2 DSLs) and usability (1 DSL).

	\item[Benchmark.] Four DSLs are evaluated based on benchmarks.
	Benchmarks are typically created for competitive evaluation in a specific domain and allow comparison of specific characteristics of the systems being evaluated\,\cite{Kistowski2015}.
	However, only for two of the four DSLs the entire benchmark suite is used, the other two selectively picked examples from benchmark suites.

	\item[Comparison.] For one DSL a systematic comparison with other DSLs is provided in which the expressiveness of the DSLs is discussed based on their syntax.
\end{description}

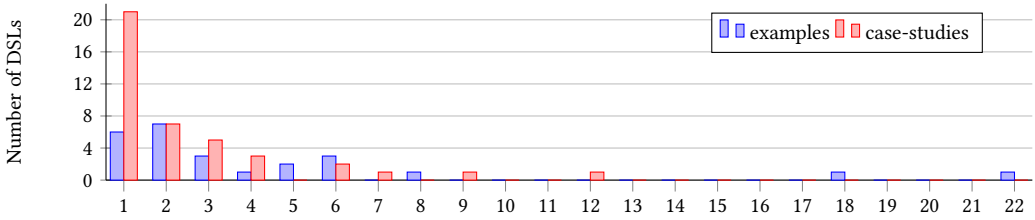
\begin{figure}[t]
	\begin{tikzpicture}
    \footnotesize
    \begin{axis}[
        ylabel=Number of DSLs,
        ytick distance=4,
        axis lines=left,
        axis line style={-},
        enlarge x limits=0.02,
        legend style={at={(0.8,0.95)},
        anchor=north, legend columns=0},
        xmajorgrids=false,
        xminorgrids=false,
        ymin=0,
        ymax=22,
        ymajorgrids=true,
        ybar=0pt,
        bar width=0.18cm,
        xtick=data,
        width=\textwidth,
        height=3.5cm,
    ]
    \addplot table [x=numberdsl, y=examples] {figures/casestudies.data};
    \addplot table [x=numberdsl, y=case-studies] {figures/casestudies.data};
    \legend{examples, case-studies}
    \end{axis}
\end{tikzpicture}
	\vspace{-10pt}
	\caption{Number of investigated case studies/examples}
	\label{fig:statistics:casestudies}
	\vspace{-12pt}
\end{figure}

\looseness=-1
Since examples and case studies are the main methods used for evaluating DSLs, in \cref{fig:statistics:casestudies}, we depict the number of case studies and examples presented for the DSLs. The maximum number is 22 for one DSL, but on average only 3.2 examples or case studies per DSL are presented. The median is two.
Altogether, the empirical evaluation of the DSLs is mostly focused on individual cases that may be considered representative, but rigorous evaluation methods are often missing.

\subsubsection{Formal Methods}
\edit{The publications of 40 DSLs (33\,\%) state some form of security guarantee.
We observed a strong correlation between the availability of an attacker model and a security guarantee:
79\,\% of the DSLs with an attacker model have a security guarantee and explicitly state the security objective.
On the other hand, DSLs with a focus on access control do not have an attacker model (81\,\%) and do not provide a security guarantee in most cases (93\,\%).
In general, when a security guarantee is provided by a DSL, it should be confirmed by some form of evaluation.
Since a significant number of DSLs have a formal proof, we provide some details on this aspect.}

\begin{description}
	\item[Not applicable.] 67\,\% of our DSLs do not come with a formal proof, since formal proof techniques are not applicable to them, e.g., due to their informal nature.
	Only 33\,\% of the DSLs provide a security guarantee, for example, many languages in the field of information flow control assert a form
	of non-interference, a concept introduced by Goguen and Meseguer\,\cite{DBLP:conf/sp/GoguenM82a},
	which implies that the observable output of a system must be independent of its sensitive input.
	\item[Formal proof.] Out of the set of 30 DSLs with a security guarantee, 73\,\% provide a formal proof that assures the correctness of the claimed guarantee. These are mostly from domains with strong formal foundations, e.g., cryptography or information flow control.
	\item[No proof.] \edit{Still, 12 DSLs remain that contain security aspects that could be accompanied by a formal proof, but no formal proof is provided. This are 10\% of all DSLs and 40\,\% of the 30 DSL for which a formal proof would be applicable.}
	
\end{description}

\subsubsection{Objectives}

After identifying the method used for evaluation, we investigated what are the objectives of the evaluations.
We identified 13 different objectives of the evaluation as well as the illustration of the DSL on examples, which is presented as an evaluation of the DSL.

\begin{description}
	\item[Applicability.] Nine DSLs are evaluated concerning their applicability to real-world systems and application scenarios, i.e., whether the assumptions of the scientific worlds meet practice. Thereby, the focus is primarily on validating that variations that might occur in practice are supported technically and not whether the proposed technique provides good results for all variations.
	\item[Developer overhead.] One DSL is evaluated concerning the overhead it introduces for developers using the DSL.
	However, this is assessed by the authors themselves in a case study they conducted, not by a user study as one might expect.
	\item[Effectiveness.] The effectiveness to actually achieve the intended security goal is evaluated for 8 DSLs. Effectiveness is evaluated via controlled experiments. On average, 9.3 more or less representative examples are used for this purpose, with the minimum being only one example and the maximum being 22 examples.
	\edit{Note that the effectiveness of a DSL is not limited to concrete security guarantees; indirect security benefits are also relevant. For example, by simplifying the representation of an instantiation, a DSL helps to detect or avoid errors.}
	\item[Expressiveness.] For 9 DSLs their expressiveness is evaluated. For this purpose, only what can be expressed using the syntax of the language is considered and not how the supporting tooling applies to real-world systems as also considered in the evaluations targeting applicability.
	\item[Feasibility.] Feasibility is usually evaluated on individual case studies by demonstrating that the proposed technique is feasible to solving the targeted problem. 
	Thereby, the main focus is to demonstrate that the target problem is solvable using the DSL in some setting, which not necessarily has to be generalizable or entirely realistic.
	Corresponding to case studies being the mainly applied evaluation method, the majority of the DSLs (40 DSLs) is evaluated concerning their principle feasibility for the intended task.
	\item[Performance.] The concrete performance of the DSLs, typically focusing on the runtime of the corresponding tooling, is the evaluated for 18 DSLs.
	\item[Precision.] For one DSL the precision of a static analysis is evaluated.
	\item[Scalability.] Four DSLs are evaluated how they scale for huge inputs.
	\item[Understandability.] Whether stakeholders can easily understand what is expressed using the DSLs is evaluated for 2 DSLs.
	\item[Usability.] For 5 DSLs the evaluation targets the concrete application of the DSLs by developers and how their experience in applying the DSL.
	\item[Usefulness.] Whether the DSLs are useful in practice is evaluated for 2 DSLs. One time in a user study and once based on examples. Taking the benefits of the DSLs and supporting tooling as granted, the focus is on their embedding into practical processes and how the engineering of secure systems is improved in relation to overhead added.
	\item[Illustrate.] The illustration of what the DSL looks like to convey the idea of how it will be applied is based on examples for 12 DSLs and is sold as a form of empirical evaluation. However, these examples are relatively small and may not fully align with the scope of a case study, which could potentially impact the validity of conclusions drawn from them.
\end{description}

\subsubsection{Practical Application}
Only a minority of 14 DSLs (12\,\%) report on experiences from applications in practice.
Three of these DSLs are the OASIS standards \textsf{SAML}, \textsf{XACML}, and \textsf{WS-SecurityPolicy} that are widely used in practice. The mature academic security frameworks \textsf{CORAS}, \textsf{STS}, and \textsf{UMLsec} have been applied in various industrial context in projects and papers extending their original publications. All others have been applied to individual open-source projects or been evaluated together with industry partners.

\begin{observation}
	Not only have most security DSLs rarely been applied in practice, their empirical validation is often lacking or of low quality. \edit{However, in some domains, a significant share of DSLs offer formal security guarantees and}
    those DSLs that have been evaluated in practice or come from open source projects show a high interest from the industry.
\end{observation}

\section{Discussion}
\label{sec:discussion}
We now discuss the challenges we identified related to the development and use of security DSLs, opportunities for future research, and a vision for integrating security DSLs throughout the SDLC.

\subsection{Towards an Evaluation Framework}
\label{sec:framework}
	\edit{
			Despite analyzing a large number of security DSLs, we found few reports of their practical usage in the scientific literature.
		Instead, we observed an unsystematic empirical evaluation of security DSLs, underscoring the need for a more systematic assessment.
		To start the construction of a general evaluation framework for assessing security DSLs,}
	\cref{tab:framework} aggregates the details on the formal and empirical evaluation of security DSLs\edit{, as well as aspects identified in the general literature on DSL quality} inspired by the Goal-Question-Metric paradigm for defining and interpreting software measurement\,\cite{Basili1992}.
	The goals were derived from the aspects addressed \edit{in RQ6 focusing on the evaluation of security DSLs (see \cref{sec:rq6-eval})}, removing aspects that are claimed to be part of an evaluation but do not come with measurable metrics or qualitative arguments, i.e., \textit{Illustrate}.
	For every goal, we extracted relevant questions to answer when assessing a security DSL as well as suitable metrics or qualitative arguments for answering the questions.
	When multiple aspects extracted from the papers containing the evaluation of a DSL were merged into one goal, the individual aspects are covered by the questions.
	\edit
	{Thereafter, we triangulated with general literature on usability aspects of DSLs and evaluating DSLs, extending goals, questions, and metrics with additional ones observed in the literature (annotated with references in \cref{tab:framework}).
	Particularly relevant was the qualitative DSL assessment framework\,\cite{Kahraman2015} and the usability evaluation taxonomy\,\cite{Rodrigues2017}.
	We assessed every aspect presented in the literature with respect to the specifics of security DSLs, considering also insights from multiple perspectives, e.g., reusability of artifacts written in a DSL is considered in a framework as its own category\,\cite{Kahraman2015}, but a user study shows that it did not contribute to the success of the DSL\,\cite{Hermans2009}.
	Finally, we performed a sanitation step, merging closely related aspects into a single aspect, e.g., considering \textit{Scalability} as a subaspect of \textit{Performance}.
}

\edit{
	The investigated literature on security DSLs has focused on technical aspects such as correctness, effectiveness, and performance, while more general works on DSL qualities do not explicitly consider correctness and how it is ensured. 
	Yet, these aspects are paramount in the domain of security.
	While the literature on security DSLs also discusses functional suitability and productivity, relevant aspects for practical use, such as usability, are barely discussed.
	Existing quality literature demonstrates the breadth of the spectrum that must be considered.
}

	\begin{table}
	\caption{Towards an evaluation framework: Goals, questions, and methods for assessing security~DSLs}
	\label{tab:framework}
	\vspace{-.5cm}
	\scriptsize
	\setlength{\tabcolsep}{3pt}
	\begin{tabularx}{\columnwidth}{p{1.3cm}Xp{1.28cm}p{3.98cm}}
		\toprule
		\textbf{Goals} 	& \textbf{Questions} & \textbf{Methods} & \textbf{Metrics/Qualita\-tive Arguments} \\    
		\midrule
		
		\textit{Compatibility*}\newline
		\textsubscript{\cite{Kahraman2015,Hermans2009}} & \textit{How compatible is the DSL with domain-specific (development) process?}		& Case Studies, Examples & SDLC phases\textsubscript{\cite{Kahraman2015}}  \\
		\cline{2-4}
		& \textit{Has the DSL the capability to operate with other elements in the domain?}						& Case Studies, Examples & artifacts supported\textsubscript{\cite{Kahraman2015}}, \textit{tailoring to specific artifacts$^*_\text{\cite{Hermans2009}}$}\\	
		\midrule
		
		Correctness			 		& Is the DSL's implementation correct in respect to the security aspect expressed? 	& Formal Proofs 			& n/a \\
		\cline{2-4}
									& How understandable is the security DSL for stakeholders? 							& User Studies 				& \#\,correctly understood concepts, user rating, semiotic clarity\\
		
		\midrule
	
		Effectiveness \textsubscript{\cite{Barisic2011,Rodrigues2017,Hermans2009}} 		& How effective is the DSL in reaching its security goal? 							& Benchmarks, Experiments 	& recall (\#\,vulnerabilites found/pre\-vent\-ed)\textsubscript{\cite{Hermans2009}}, precision (\# false positives) \\
		\cline{3-4}
							&																					& Case Studies				& \#\,vulnerabilites found/prevented \\
		\cline{2-4}
							& How effective is the DSL in facilitating security engineering tasks?				& Experiment, Case Studies	& \# changes needed, \# LoC needed/saved\\
		\cline{2-4}
							& How good is the quality reached by the security DSL?								& Benchmarks, Experiments	&  size of output (i.e., LoC, \#\,gates, chip area, power consumption)\\
		
		\midrule
		
		\textit{Extensibility*}\newline \textsubscript{\cite{Kahraman2015,Barisic2012}} & \textit{Has the DSL mechanisms to add user-specific features?} & \textit{Case Study, Examples} & \textit{extension points$^*_\text{\cite{Hermans2009}}$}\\
		
		\midrule
		
		Functional Suitability \textsubscript{\cite{Kahraman2015,Hermans2009}} 		& Can all relevant scenarios be expressed in the security DSL, i.e., is it complete\textsubscript{\cite{Kahraman2015,Hermans2009}}?  						& Examples, Case Studies	& list of concepts expressible\textsubscript{\cite{Hermans2009}}, \textit{number of concepts$^*_\text{\cite{Hermans2009}}$}, \textit{re\-stric\-tions$^*_\text{\cite{Hermans2009}}$}\\								
		\cline{2-4}
					 		& Is the DSL appropriate to solve the intended problem?								& Examples, Case Studies	& problem solved\\
		\cline{2-4}
					 		& Is the security DSL applicable to real-world systems and scenarios? 				& Case Studies 				& practical constrains met, barriers to adoption\\
		
		\midrule
		
		\textit{Integrability*}\newline \textsubscript{\cite{Kahraman2015}} & \textit{How well can the DSL be integrated with other languages used?} & \textit{Case Studies, Examples} & \textit{supported languages$^*_\text{\cite{Kahraman2015}}$}, 
																																									\textit{reusable elements$^*_\text{\cite{Hermans2009}}$} \\
		\midrule
		Performance 		& How much computation resources are needed by the tooling of the DSL?				& Benchmarks, Experiments	& execution time\textsubscript{\cite{Mohagheghi2010}}, memory usage, storage usage \\
		\cline{2-4}
							& How does the security DSL impact the runtime performance of the software system?$_\text{\cite{Mohagheghi2010}}$							& Benchmarks, Experiments	& throughput, delay \\
		\cline{2-4}
							& How does the runtime performance scale?											& Benchmarks, Experiments	& scalability for large/complex cases \\

		\midrule
		
		Productivity \textsubscript{\cite{Kahraman2015,Rodrigues2017,Barisic2012,Hermans2009}} 		& How much overhead for developers does the security DSL introduce?					& User Studies, User Surveys 				&
					LLOC written, 
					\textit{mental effort$^*_\text{\cite{Barisic2011}}$}
					\textit{number of ac\-tiv\-ities$^*_\text{\cite{Kahraman2015}}$}, 
					\textit{mouse} \textit{move\-ments$^*_{\text{\cite{Barisic2012}}}$}, 
					\textit{keystrokes$^*_{\text{\cite{Barisic2012}}}$} 
					\\
		\cline{2-4}
		& How does using the security DSL impact development time?							& User Studies				& time spent$_\text{\cite{Kahraman2015,Barisic2012,Barisic2011,Ingibergsson2018}}$\\
		\cline{2-4}
		& How does the resource usage change to develop a secure system?					& User Studies				& \textit{human resources$^*_\text{\cite{Kahraman2015}}$}, code generated$_\text{\cite{Hermans2009}}$\\
		
		\midrule
		
		Usability\newline \textsubscript{\cite{Kahraman2015,Barisic2011,Barisic2012}}
		\textsubscript{\cite{Rodrigues2017,Hermans2009}}			& How usable is the syntax of the security DSL?$_\text{\cite{Kahraman2015}}$		& User Studies 				& time to learn\textsubscript{\cite{Hermans2009}}, \textit{time to under\-stand$^*_\text{\cite{Mohagheghi2010}}$}, number of mistakes$_{\text{\cite{Barisic2012,Ingibergsson2018}}}$ \\
													&																			& Examples					& \textit{operable language elements$^*_\text{\cite{Kahraman2015}}$} \\

		\cline{3-4}
													&	& User Surveys				& 	\textit{accessibility$^*_\text{\cite{Barisic2011}}$}, 
															\textit{attractiveness$^*_\text{\cite{Kahraman2015}}$}, 
																						\textit{com\-prehensibility$^*_\text{\cite{Kahraman2015}}$}, 
																						\textit{intuitive\-ness$^*_{\text{\cite{Rodrigues2017}}}$}, 
																						\textit{perceived complexity$^*_{\text{\cite{Rodrigues2017}}}$}, 
																						\textit{satisfaction$^*_{\text{\cite{Barisic2012,Barisic2011}}}$}
																						\\
&																			& Examples					& \textit{operable language elements$^*_\text{\cite{Kahraman2015}}$} \\
		\cline{2-4}
													& How usable is the tooling  corresponding to the security DSL? & User Studies & user rating, \textit{number of activities$^*_\text{\cite{Kahraman2015}}$}, \textit{quality of documentation$^*_\text{\cite{Kahraman2015}}$} \\ 
		\bottomrule
	\end{tabularx}
	\vspace{-.2cm}
	\textit{* Derived from additional literature and not observed in the papers/material on the security DSLs}\hfill
	\vspace{-.4cm}
\end{table}

\subsection{Challenges \& Opportunities}
\looseness=-1
\edit{
Although security DSLs clearly contribute to the security of a system, the low number of documented practical evaluations indicate low rate of adoption.}
Investigating the barriers to practical adoption of security DSLs represents a significant research opportunity.

\looseness=-1
Barriers to adoption that we have identified are primarily a lack of availability and usability.
For two-thirds of the DSLs, the framework presented in the publications is not publicly available.
The prevalent primitive level of usability, e.g., due to missing editing support for the DSL, is certainly a limitation for practical adoption.
We assume that the focus on novelty and theoretical foundations in academia in combination with a lack of incentives for practicality and maintenance are the cause for this situation.
Besides this, particularly design-time security DSLs require specific artifacts, e.g., models, that are often not available in practice\,\cite{Gorschek2014}.
However, industrial threat modeling frameworks such as Microsoft's STRIDE significantly rely on design models, i.e, data flow diagrams\,\cite{Shostack2008}.
Nevertheless, a few security DSLs have led to commercial products, enjoyed widespread adoption beyond the research community, or become an international standard.
However, effective editing support is needed to put the DSLs into practice, and involves relevant research opportunities, particularly for the human-based security community, such as identifying what developers need to effectively use a security DSL.
In line with this, an initial study focusing on UMLsec\,\cite{Ebad2022} shows that developers consider the prototypical state of the tooling, i.e., leading to a limited user experience, and the required training in using the DSL to be main reasons for the lack of adoption.

An objective, standardized evaluation framework would help to compare security DSLs, e.g., with respect to their strength and weaknesses, and assist practitioners in identifying the most appropriate one for their needs. 
Our initial framework for assessing security DSLs (cf. \cref{tab:framework}) and our supplementary material\,\cite{Krausz2025}, consisting of detailed labeling per DSL, could be used as a basis for developing this framework. As the categories of security DSLs identified in this work address different domains and problems, a framework tailored to each category may be required. In this regard, as related work (see \cref{sec:related}), we discuss existing surveys that address individual categories.

Our literature review shows that DSLs exist for nearly all major security aspects and system components, leaving few gaps.
Access control DSLs, such as SAML, WS-SecurityPolicy, and XACML, are widely used and considered mature, with limited need for further extensions.
DSLs for information flow control and secure computation increasingly target specialized contexts, such as concurrency, distributed systems, or MPC.
\edit{
	However, despite the importance of testing systems for security, we did not observe any DSLs specific to writing security tests or the secure deployment of systems.
While there are many general DSLs for specifying test cases, such as Robot\,\cite{Bisht2013}, and even sub-DSLs within it\,\cite{Peldszus2023}, security-focused DSLs appear to be absent. 
Perhaps existing DSLs are sufficient for specifying security tests and require further automation to generate them from other DSLs, for example.
Conversely, current studies show that security testing is still primarily manual,\,\cite{Hermann2025icse}, e.g., based on pen testing, which might indicate a lack of need for security-specific testing DSLs.
}

Although most DSLs have been published in security venues, the security community itself is fragmented into several subdomains.
The isolation of each DSL ultimately reflects this fragmentation.
Unfortunately, we observe a low level of integration and combination.
Most of DSLs produce only single artifacts and are not integrated into other frameworks.
Nevertheless, there are some security DSLs that cover multiple phases of the SDLC and demonstrate the benefits of doing so, i.e., ensuring that planned security requirements are addressed as planned in later development phases.

Particularly evident is, for scientific standards, the low level of evaluation for the DSLs we discovered.
While the languages in the cryptography and information flow control domains usually provide strong formal security guarantees to undermine their effectiveness, most DSLs have not been evaluated with clear methodology.
Many papers only demonstrate the feasibility of their language with small examples, while user studies have only been executed for a few DSLs.
This provides an opportunity for empirical studies of the effectiveness of different security DSLs and to identify factors that influence the applicability or security gain of security DSLs.

In practice, one of the main challenges in developing secure systems is the planning of appropriate security features and their secure realization\,\cite{Hermann2025emse}.
While some categories of security DSLs usually come with tool support for generating secure code or compiling instances into executable code, these are primarily DSLs targeting custom cryptographic security features or enriching code with information flow control policies.
However, both still require developers to manually embed the security features into a system.
Although security DSLs have been used for code generation\,\cite{Arogundade2020,Peldszus2022}, these approaches mainly generate structural code around security features rather than securely embedding the features themselves.
Therefore, it only prevents a security feature from being missing altogether.
Emerging technologies, such as generative AI, have been used to automate code generation, but have security issues\,\cite{Pearce2022}.
Using DSLs as input to generative AI could be an effective building block for generating secure code, including the secure use of security features.

\subsection{Vision of an Integration Scenario}
\label{sec:scenario}
\looseness=-1
During our study, we identified various best practices and ideas that can help to effectively utilize security DSLs throughout software development.
To demonstrate how several of the DSLs examined in this work can be integrated to develop a secure system, we designed a simple example of an EHR subsystem as motivated in the Introduction.
Specifically, we mapped the DSLs to their respective phases in the SDLC, and we then highlighted how the DSLs help to address specific security issues in order to outline potential integrations.
This example consists of a measurement device (a computed tomography (CT) scanner) that sends its data over an open network to a server.

\begin{figure}[b]
	\vspace{-8pt}
	\includegraphics[width=.7\textwidth]{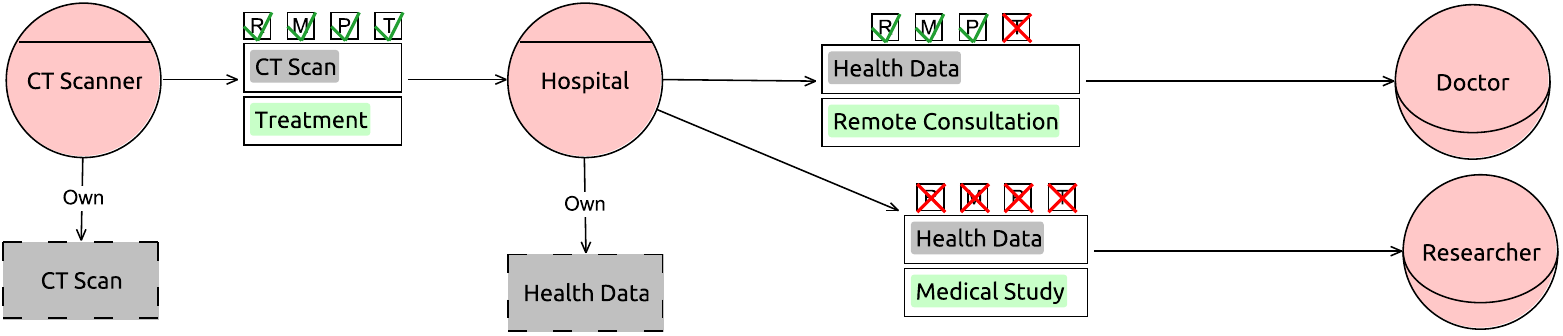}
	\vspace{-8pt}
	\caption{Authorization view for an electronic health care system in \textsf{STS}}
	\label{fig:example:sts-auth}
	\vspace{-.4cm}
\end{figure}

\looseness=-1
At the beginning of the development process, (security) requirements are elicited and specified with \textsf{STS} and its supporting tools.
The functional requirements of the system described above are modeled using a goal model notation oriented on SI* (see \cref{fig:example:sts-scocial}).
Thereby, data planned to be exchanged is already explicitly specified on an abstract level.
Here, the sensitivity of the measurement data gets identified and documented, for example, the CT images are labeled as security-critical data.
Also using \textsf{STS}, basic authorization requirements for the intended usage scenarios are planned (see \cref{fig:example:sts-auth}).
Thereby, checks provided as part of editing support help in reasoning about the security requirements.
For example, STS allows to propagate threats through the planned system to help identifying possible impacts.
In \cref{fig:example:sts-scocial}, the tool highlights in red that the threat of the CT scanner being not reliable can propagate the treatment of a patient and might impact it.

The next step is to plan the design of the system, where proper consideration of all security requirements from above is essential, but usually difficult to achieve manually.
However, an integration between DSLs focusing on security requirements and DSLs focusing on secure system design would allow to automatically consider all security requirements.
For example, one part of designing a secure system comprises the deployment of the system in the hospital, which in our scenario is planned using UML deployment diagrams and \textsf{UMLsec}.
Thereby, security requirements initially specified in \textsf{STS} are transferred into \textsf{UMLsec} (for a practical realization of this step cf. Ahmadian et al. \cite{DBLP:conf/sigsoft/AhmadianPRJ17} and Peldszus et al. \cite{Peldszus2020b}).
In \textsf{UMLsec}, these requirements are then used to reason about the security requirements of the communication paths between the individual devices.
Since private data is sent from the device to the server, this necessitates an encrypted connection between the two units to achieve the confidentiality requirements (\mbox{$\ll$$secrecy$$\gg$} in \cref{fig:example:umlsec-deploy}) of the data.

After storing the data on the server, it should not be accessible to everyone and the authentication requirements shown in \cref{fig:example:sts-auth} must be realized.
To this end, proper access control constraints must be specified so that,
e.g., a physician can read the sensitive measurement values, but not the system administrator.
Comparable to \textsf{UMLsec}, the early authorization requirements specified in \textsf{STS} should be taken as an input and be detailed using the more fine-grained access control DSLs.
While we are not aware of any automation of the initialization of an access control DSL, DSLs such as \textsf{UMLsec} or \textsf{SecureUML} allow to immediately annotate design models with access control constraints, therefore supporting integrated development of more fine grained access control systems.
In the best case, specification are created that can be transformed into a format that can be interpreted at runtime by an access control engine to avoid manual implementation of the access control policies.

\looseness=-1
To prevent unintended information leakage at a lower level, \textsf{Jif} is used to implement data gathering logic.
For example, a measurement data variable can be declared in \textsf{Jif} as shown in \cref{lst:jif}.
The security label in the curly brackets assures, that data generated by the scanner can only be read by the scanner itself and any entity that can act as \texttt{AuthorizedUser}.
Furthermore, the measurement can only be modified by the scanner to ensure its integrity.
This code can be compiled to Java and thus seamlessly integrated with code written in Java.
However, as for the design models, the data flow policies expressed in \textsf{Jif} must comply to the data flows planned during requirements engineering (cf. \cref{fig:example:sts-scocial,fig:example:sts-auth}) as well as all further refinements, e.g, using security DSLs at design time.
While we are not aware of any compliance check between data flow policies in Jif and ones expressed in other security DSLs, there are works that check whether design-time data flow contracts are properly implemented using static or dynamic analysis or by tailoring taint analyzers \cite{Tuma2023,Peldszus2024,Lanzinger2024}

\begin{lstlisting}[basicstyle=\scriptsize, float,label={lst:jif},belowskip=-3em,
	caption=Excerpt from a data class with secure data flow constraints written in \textsc{Jif},
	language=Java,
	belowskip=-2.5 \baselineskip]
public class CTScan {
	// raw scan with data flow policy
	byte[]{Scanner->AuthorizedUser; Scanner<-*} measurement;
}
\end{lstlisting}

\looseness=-1
To allow participating in medical studies across multiple institutes, while avoiding disclosure of patient measurements due to privacy regulations, secure multi party computation is used to handle statistical requests, e,g., determining median values of patient measurements.
This can be implemented in \textsf{SFDL 2.0} with little cryptographic knowledge and FairplayMP generates code, therefore allowing implementation-level integration.
%
The existence of several information-flow-sensitive DSLs, such as \textsf{Jif} or \textsf{SFDL 2.0}, points to the final challenge in the outlined integration, which is to decide on the appropriate DSLs to select for fulfilling the individual security requirements.
Here, the overview provided in this paper helps to manually select suitable security DSLs for individual tasks.

\edit{\looseness=-1
	Note that we selected the DSLs so that integration and combination are already feasible or would only require reasonable effort.
	In general, combining multiple DSLs, can lead to significant complexity due to semantic\,\cite{Neto2013} and technical\,\cite{Horst2013} differences. 
	Integrability and compatibility are also featured as quality aspects of our proposed evaluation framework.
	For integrating independent components, different types of reuse need to be considered, such as type, implementation, and binding hierarchies\,\cite{Mezini2002}.
	Composing multiple DSLs requires creating interfaces for composing term representations and defining interpretations across languages\,\cite{Hofer2010}.
	The necessary steps for systematically integrating DSLs and their common concepts are open research questions\,\cite{Heithoff2023,Pfeiffer2023}. 
}
\section{Threats to Validity}
\label{sec:threats}
Our systematic methodology contributes to our confidence that we have identified all security domains where a significant number of DSLs is published, and that we have included its most influential DSLs. Still, any empirical study is prone to threats to validity\,\cite{Jedlitschka2008RES}.
\looseness=-1
A potential threat is that our methodology may have missed relevant domains, especially domains that do not match our search terms and are not referenced by the papers we included.
To mitigate this, we employed snowballing as recommended in ACM SIGSOFT's Empirical Standards\,\cite{Ralph2021}.
Since all DSLs known to us from our research experience, including lesser-known DSLs, were covered, this indicates that the employed methodology, search terms, and filter criteria were well chosen.
\edit{
	We only included DSLs reported in peer-reviewed academic papers to ensure quality.
	However, we also considered other sources, such as project pages and wikis related to the DSLs. This ensured an informed perspective for our analysis.
}
\looseness=-1

There might be a bias due to the research background of the authors conducting the study, which could have impacted their execution, e.g., decisions on what is considered as a security DSL. To mitigate this threat, two authors with different backgrounds conducted each step of the investigation.
Still, the authors could have misunderstood the papers and done incorrect classifications. To mitigate this threat, the authors actively discussed ambiguous classifications
In addition, when analyzing the results, the authors actively read papers where the extracted statements and corresponding classifications did not seem to match. Only minor discrepancies were identified and fixed.

The papers analyzed may not be representative of security DSLs applied in practice or of all DSLs considered in the security community.
However, the significant number of papers and DSLs examined reduces this threat.
Furthermore, they stem from many research communities, covering scientific and practitioner-driven DSLs.
\edit{
	During our analysis, we observed that DSLs targeting the testing or deployment phases of the software development lifecycle were underrepresented in our sample.
	Therefore, we refrain from drawing conclusions specific to these SDLC phases and instead openly discuss possible reasons for this observation.
}

In practice, DSLs may be more deeply integrated than observed, while not yet investigated by researchers.
This could be a threat to our conclusions---but very unlikely, since our statistics show that most DSLs are relatively old, while we also covered newer works.
\section{Related Work} \label{sec:related}

\looseness=-1
Only few works systematically examine security-related DSLs, mostly with a narrow scope.
Two works\,\cite{do2012systematic, DBLP:journals/infsof/KosarBM16}
provide an overview of the domains where DSLs have been used, but are not security-specific.
On the other hand, multiple surveys review tools, frameworks, and DSLs for single security subdomains.
Tools and DSLs for the development, specification, and implementation of cryptography are evaluated
in the context of computer-aided cryptography\,\cite{DBLP:conf/sp/BarbosaBBBCLP21}.
Another area that induced many new DSLs in the last years, are smart contract languages. This area is surveyed by two recent
works\,\cite{DBLP:journals/csur/Varela-VacaQ21, DBLP:journals/jss/VaccaSVC21}.
A subset of the DSLs that we cover
are included in publications specifically about policy specification
languages\,\cite{DBLP:journals/cn/HanL12, DBLP:journals/corr/Kasem-MadaniM15, leicht2019survey}.
Also, DSLs from the domain of security-aware system modeling are surveyed\,\cite{DBLP:conf/apsec/HachemPCBA16}.
Lallie et al.\,\cite{DBLP:journals/csr/LallieDB20} review the visual syntax of attack graphs.
Finally, four surveys\,\cite{Uzunov2012, Nguyen2015, DenBerghe2017, Mashkoor2023} on model-driven security engineering mention various design-time security DSLs, but mainly focus on model-based security engineering processes. A systematic overview and comparison, as in our work, is not provided.

\section{Conclusion} \label{sec:conclusion}
\looseness=-1
We presented the first holistic survey of 120 DSLs for security.
Our focus were DSLs presented in the peer-reviewed academic literature to ensure quality and transparency.
We learned that the security research community proposed security DSL for nearly every security aspect.
\edit{Still, not all phases of the software development lifecycle are equally addressed.}
Most DSLs, address the design or implementation phases by enhancing the expression of design models, policies or code, \edit{while deployment and testing is not explicitly covered.
As many DSLs specific to the underrepresented phases exist, e.g., for specifying test cases \cite{Bisht2013,Peldszus2023}, one reason could be that security-specific DSLs are not necessary to specify security tests.}
We conclude that the integration, combination, usability, and evaluation of security-oriented DSLs still offers many opportunities for further research.

\looseness=-1
Future work comprises expanding the initial framework of criteria for comparing and evaluating security DSLs.
To this end, identifying reasons for the lack of security DSL adoption and derive metrics to evaluate their potential use in practice would be a valuable contribution.
However, this requires a study with practitioners, which is outside the scope of this work.
Finally, recall that generative AI and coding assistants have substantially enhanced software development, but do not provide guarantees about the correctness of code, let alone about the security of it \cite{Pearce2022,Perry2023}. The security DSLs we surveyed provide strong guarantees when used correctly. Finding modalities and methods of integrating the DSLs with coding assistants would be promising future research directions.


\bibliographystyle{ACM-Reference-Format}
\bibliography{bibliography, slr}


\end{document}